\documentclass[3p]{elsarticle}

\usepackage{amssymb,bm}
\usepackage{mathrsfs,amsmath}
\usepackage{amsfonts}
\usepackage{subfigure}
\usepackage{color}
\usepackage{graphicx}
\usepackage{tabu}
\usepackage{undertilde}
\usepackage{lineno,hyperref}
\usepackage{multirow}
\usepackage{enumitem}
\newcommand{\captionabove}[2][]{%
    \vskip-\abovecaptionskip
    \vskip+\belowcaptionskip
    \ifx\@nnil#1\@nnil
        \caption{#2}%
        
    \else
        \caption[#1]{#2}%
    \fi
    \vskip+\abovecaptionskip
    \vskip-\belowcaptionskip
}

\usepackage{amsmath}

\numberwithin{equation}{section}
\newcommand{\xoverbrace}[2][\vphantom{\dfrac{A}{A}}]{\overbrace{#1#2}}

\makeatletter
\def\ps@pprintTitle{%
	\let\@oddhead\@empty
	\let\@evenhead\@empty
	\def\@oddfoot{}%
	\let\@evenfoot\@oddfoot}
\makeatother

\begin{document}

\begin{frontmatter}

\title{Deep material network with cohesive layers: Multi-stage training and interfacial failure analysis}
\address[Zeliangaddress]{Livermore Software Technology, an ANSYS company, Livermore, CA 94551, USA}
\cortext[mycorrespondingauthor]{Corresponding author.}
\author[Zeliangaddress]{Zeliang Liu\corref{mycorrespondingauthor}}
\ead{zeliang.liu@ansys.com}
\begin{abstract}
A fundamental issue in multiscale materials modeling and design is the consideration of traction-separation behavior at the interface. {By enriching the deep material network (DMN) with cohesive layers, the paper presents a novel data-driven material model which enables accurate and efficient prediction of multiscale responses for heterogeneous materials with interfacial effect.} In the newly invoked cohesive building block, the fitting parameters have physical meanings related to the length scale and orientation of the cohesive layer. It is shown that the enriched material network can be effectively optimized via a multi-stage training strategy, with training data generated only from linear elastic direct numerical simulation (DNS). The extrapolation capability of the method to unknown material and loading spaces is demonstrated through the debonding analysis of a unidirectional fiber-reinforced composite, where the interface behavior is governed by an irreversible softening mixed-mode cohesive law. Its predictive accuracy is validated against the nonlinear path-dependent DNS results, and the reduction in computational time is particularly significant.
\end{abstract}
\end{frontmatter}
\section{Introduction}
Materials have deformable interfaces across fine-scale phases, and an emerging need in multiscale materials modeling and design is to capture the effects of interfaces at various length and time scales. Some representative examples are fiber-matrix debonding in carbon fiber reinforced polymer (CFRP) composites \cite{melro2013micromechanical}, Mullins effect due to stress softening at the particle/matrix interface \cite{cantournet2009mullins}, void nucleation in ductile fracture of metals \cite{vernerey20063,mcveigh2007interactive}, and grain-boundary effects in polycrystalline materials \cite{simone2006generalized,espinosa2003grain}. One can see that the interface, in many cases, is the critical region involving damage initiation and evolution, thus, it plays an important role in the failure analysis of modern material systems. On the other hand, as the interface is a geometric object with one dimension less than the bulk material, it also results in various interesting phenomena related to the size effect of microstructures \cite{needleman1987continuum,duan2005size}.

A plethora of multiscale material models have been developed to consider the interfacial effect. However, empirical models are problem-dependent and may lose the essential physics of the interface. Analytical micromechanics methods based on the Eshelby's solution \cite{eshelby1957determination} and mean-field theories \cite{mori1973average,hill1965self} can be enhanced to incorporate the interfacial effect, but they are usually limited to regular geometries and weakly nonlinear behaviors \cite{qu1993effect,tan2005mori,duan2005size}. To resolve these issues, computational homogenization is emerging as an accurate way of deriving multiscale material responses from the representative volume element (RVE) \cite{geers2010multi}, where the microstructures are explicitly constructed and discretized by numerical models. 

In recent years, direct numerical simulation (DNS) of RVE has been widely used to understand the role of interface in heterogeneous materials.  To describe the traction-separation relationships across the interfaces, various cohesive laws have been developed \cite{park2011cohesive} and further built into DNS tools with cohesive-element formulations, such as the finite element method (FEM) \cite{camacho1996computational,ortiz1999finite}, the extended finite element method (XFEM) \cite{hettich2006interface,belytschko2009review}, the generalized finite element method (GFEM) \cite{simone2006generalized}, and isogeometric analysis (IGA) \cite{deng2015isogeometric,bazilevs2018new}. The pioneer work done by Needleman \cite{needleman1987continuum} simulated a periodic array of rigid spherical inclusions to investigate the cause of void nucleation due to the inclusion debonding. Melro et al. \cite{melro2013micromechanical} developed RVE models for polymer composite reinforced by unidirectional fibers and evaluated the influence of fiber-matrix interface on the strength properties of the composite. Zhao et al. \cite{zhao2017simulation} utilized the XFEM to simulate the multi-void nucleation process due to interface debonding under various loading conditions and particle size/shape distributions. However, limited by the large computational cost of DNS, these full-field RVE models were mainly used to validate and calibrate macroscopic empirical material models.

In the perspective of integrated multiscale frameworks like concurrent simulation and materials design \cite{liu2018microstructural,liu2019transfer}, the efficiency of RVE analysis arises as a fundamental issue. Many reduced-order methods have been proposed for fast RVE homogenization with interfacial failure, such as the Voronoi cell method \cite{ghosh2000interfacial,ghosh2007concurrent}, eigendeformation-based method\cite{oskay2007a,yuan2009multiple,zhang2015eigenstrain}, self-consistent clustering analysis \cite{liu2016self,liu2018microstructural,liu2018data,shakoor2017data,li2019clustering}, and proper orthogonal decomposition (POD) \cite{oliver2017reduced}. Specifically, the SCA and POD-based methods are data-driven since their reduced bases are extracted from mechanical data in the offline training stage. {Feed-forward neural network and recurrent neural network have also been used to fit the stress-strain responses of elastic and history-dependent materials \cite{ghaboussi1991knowledge,le2015computational,bessa2017,wang2018multiscale,li2019clustering}. Regarding interface modeling, Wang et al.\cite{wang2019meta} recently proposed a deep reinforcement learning framework based on directed graph for deriving the traction-separation law.} As the training data is usually limited due to the expense of physical or numerical experiments, a key challenge faced by most data-driven methods is the credibility of extrapolation to unknown material and loading spaces, especially when nonlinear softening interfaces are involved.

An alternative way of mining the mechanical data is to learn the hidden topology representation of the RVE. Following this idea, Liu et al. \cite{liu2019deep,liu2019exploring} proposed the deep material network (DMN) for data-driven material modeling, which describes the RVE by a network structure based on physics-based building blocks. Essentially, all fitting parameters of DMN are interpretable with physical meanings related to the RVE morphology. Trained only by linear elastic data, DMN can be extrapolated to capture highly nonlinear material behaviors for various applications, including hyperelastic rubber composite under large deformation, polycrystalline materials with rate-dependent crystal plasticity, and various types of CFRPs. Recently, a transfer-learning framework of DMN is further introduced to generate a unified set of material databases covering the full-range structure-property relationship \cite{liu2019transfer}.

This paper extends the deep material network framework \cite{liu2019exploring} to 3-dimensional heterogeneous materials with deformable interfaces. The main contributions of the work are:
\begin{itemize}[noitemsep]
\item New architecture of material network enriched by adaptable cohesive networks;
\item Physics-based cohesive building block where a reciprocal length parameter is introduced to describe the size effect of interface;
\item Multi-stage training strategy to reduce the number of fitting parameters in the machine-learning process at each stage;
\item Extrapolation of the enriched DMN from linear elasticity to interfacial failure analysis with irreversible softening cohesive laws.
\end{itemize}

The remainder of this paper is organized as follows. In Section \ref{sec:arch}, the new design of DMN architecture with cohesive layers is discussed in detail. Analytical solutions of the cohesive building block are derived to provide the basis of data propagation in the network. Section \ref{sec:multistage} presents the multi-stage training strategy based on linear elastic offline DNS data. Section \ref{sec:coh} briefly introduces the mix-mode cohesive law and necessary viscous regularization for implicit analysis in the online stage. Section \ref{sec:application} shows the application of DMN to the interfacial debonding analysis of a unidirectional fiber-reinforced composite, and another example on a particle-reinforced composite is also provided in \ref{ap:ap3}.  Concluding remarks and future work are given in Section \ref{sec:conclusion}.

\section{New DMN architecture with enrichment}\label{sec:arch}
\begin{figure}[!t]
	\centering
	\graphicspath{{Figures/}}
	\includegraphics[clip=true,trim = 5.0cm 4cm 5cm 4.0cm, width = 0.8\textwidth]{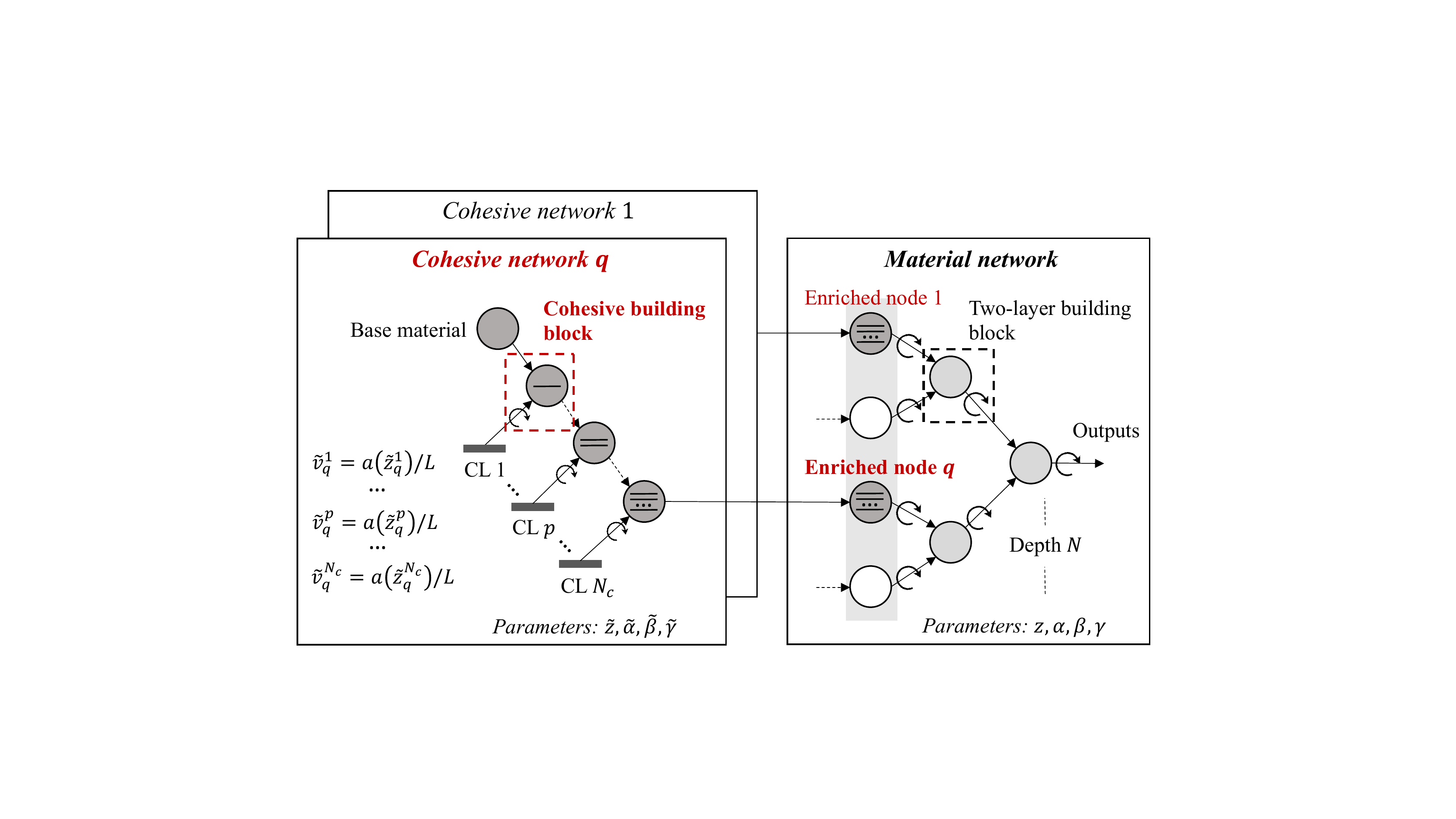}
	\caption{{Proposed architecture of deep material network enriched by cohesive networks. The depth of material network is $N$. The number of cohesive layers in each cohesive network is $N_c$. The reciprocal length parameter of the $p$-th cohesive layer in cohesive network $q$ is denoted by $\tilde{v}^p_q$, and $L$ is the characteristic length of the microstructure predefined for the better of training (see Section \ref{sec:activation}).}}
	\label{fig:framework}
\end{figure}

As a starting point, consider a material network based on a binary tree \footnote{{Current DMN architectures are based on the tree graph, in which the physical information of any node (e.g. the stiffness tensor and weight) are collected by only one upper-layer node during forward propagation. For other types of graphs \cite{wang2019meta,banerjee2019graph}, extra schemes are needed to take care how the information are distributed to multiple nodes in the upper layer.}} with a two-layer structure as the building block \cite{liu2019deep,liu2019exploring}, as shown on the right of Figure \ref{fig:framework}. The microscale constituents are input to the bottom layer, and the mechanical information such as the stiffness tensor and the residual stress is feed forward through the input-output functions defined by the physics-based building block. The macroscale material properties are extracted from the top node of the network. Let $N$ be the depth of material network, so that there are in total $2^{N-1}$ nodes at the bottom layer ($i=N$). 

The basic hypothesis of DMN is that the original 3D RVE can be represented by patterns of the two-layer structures with unknown phase fractions and block orientations \cite{liu2019exploring}. The fitting parameters of a material network are the activation $z$ of each bottom-layer node, and three rotation angles $(\alpha,\beta,\gamma)$ at each building block:
\begin{equation}
{z}^j,{\alpha}_i^k,{\beta}_i^k,{\gamma}_i^k\quad\text{with } i = 1, 2, ..., N;\:j = 1, 2, ..., 2^{N-1};\: k=1,2,...,2^{i-1}.
\end{equation} 
The activation $z^j$ determines the weight of each node in the material network. For the $k$-th node in the $i$-th layer, its weight $w^k_i$ can be expressed as a summation of activations of its descendant nodes in the bottom layer:
\begin{equation}\label{eq:wsum}
w_i^k=\sum_{j=2^{N-i}(k-1)+1}^{2^{N-i}k}w_N^j = \sum_{j=2^{N-i}(k-1)+1}^{2^{N-i}k}a(z^j).
\end{equation}
where $a(\cdot)$ is the the rectified linear unit (ReLU). Each building block takes the materials at two child nodes as inputs, and their weights determine the phase fraction of the two-layer structure. A homogenization operation is first performed to mix the two materials. The new homogenized material then undergoes a rotation defined by the angles $(\alpha,\beta,\gamma)$, and the rotated material is output to the mother node. All fitting parameters of DMN have physical meanings related to the microstructural topology, and they can be optimized through gradient-based methods by virtue of the existence of analytical solutions.

\subsection{Design of cohesive networks}
The original architecture of material network is created for heterogeneous materials with perfectly bonded phases.  Its extension to a material with interfacial effect (e.g. debonding) is limited by the design of the building block, where the two input materials are perfectly bonded.  In this regard, it is possible to consider the interface in the original two-layer building block to be deformable. However, this naive approach may suffer from two issues: 1) the interface orientation is predetermined by the building block, thus, loses the adaptivity; and 2) the physical meaning of a deformable interface in the upper part of the network ($i<N$) is not clear since the input materials of the corresponding building block usually represent a mixture of the microscale constituents.

An alternative viewpoint regards the nodes at the bottom layer of the original network as degrees of freedom (DOF). In the online computation, each active bottom-layer node serves as an independent material point, which carries the pure information of a certain microscale constituent. This provides the physical basis to introduce the interfaces only at the bottom layer.  In the paper, it is proposed to enrich each selected bottom-layer node with a so-called ``cohesive network", as shown in Figure \ref{fig:framework}.  The index of the enriched node and its corresponding cohesive network is $q$, and the total number of enriched nodes is denoted by $N_e$. A mapping function $\mathcal{J}(q)$ finds the index of enriched node $q$ in the bottom layer of the original material network:
\begin{equation}\label{eq:jq}
j = \mathcal{J}(q)\quad\text{with } q = 1, 2,...,N_e.
\end{equation}
For example, if the second cohesive network ($q=2$) enriches the third node in the bottom layer, we have $\mathcal{J}(2)=3$. 

Note that only active nodes in the bottom layer (with $z^j>0$) are necessary to be enriched. Moreover, for a two-phase material with only one type of interface, it is usually sufficient to enrich the nodes for one single phase. For example, in the material network for unidirectional (UD) fiber-reinforced composite with deformable fiber-matrix interface to be tested in Section \ref{sec:application}, only the active nodes of the fiber phase are linked to the cohesive networks. This design will help to reduce the redundancy in the machine learning model. 

In the cohesive network of the enriched node $q$, the base material should represent the same microscale constituent as the one previously at bottom layer of the original material network. As shown in Figure \ref{fig:framework}, the cohesive layers (CL), indexed by $p$, are added to the base material sequentially through the ``cohesive building block" (see Section \ref{sec:buildingblock}). For simplicity, all the cohesive networks are assumed to have the same number of cohesive layers $N_c$. Similar to the material network, the cohesive network has its own parameters, which needs to be optimized with respect to high-fidelity RVE data using gradient-based learning algorithms. The parameters of cohesive network $q$ are
\begin{equation}
\tilde{z}_q^p,\tilde{\alpha}_q^p,\tilde{\beta}_q^p,\tilde{\gamma}_q^p\quad\text{with } p = 1, 2, ..., N_c.
\end{equation} 
In this paper, the tilde symbol is used to indicate that the corresponding quantity is defined within the cohesive network. The activation of the $p$-th cohesive layer is $\tilde{z}_q^p$, and rotation angles of the $p$-th cohesive network are $\tilde{\alpha}_q^p,\tilde{\beta}_q^p,\tilde{\gamma}_q^p$. Physical meanings of all these fitting parameters will be explained in the following two subsections. Evidently, an important condition that the cohesive network should satisfy is that if all the cohesive layers are deactivated, the base material should be returned to the enriched node without any change.

The total number of parameters in the new material network with enriching cohesive networks is
\begin{equation}
N^{total} = \xoverbrace{\left(7\times 2^{N-1}-3\right)}^\text{Material network} + \xoverbrace{N_e\times N_c}^\text{Cohesive networks}.
\end{equation}
In the multi-stage training strategy to be discussed in Section \ref{sec:multistage}, not all parameters are fitted at the same time. The basic idea is to train the original material network first and then the fitted parameters ($z,\alpha,\beta,\gamma$) are treated as constants in the subsequent training of the integrated network with cohesive layers. To ease the offline sampling and training processes, the fitting parameters are optimized based on linear elastic DNS data, which is sufficient for capturing essential physical interactions among the microscale material phases and interfaces. The learned model will be extrapolated to unknown material and loading spaces, such as plasticity, interfacial failure (debonding), and loading-unloading paths.

\subsection{Activation of cohesive layers}\label{sec:activation}
\begin{figure}[!t]
	\centering
	\graphicspath{{Figures/}}
	\includegraphics[clip=true,trim = 6cm 7.0cm 6cm 7cm, width = 0.8\textwidth]{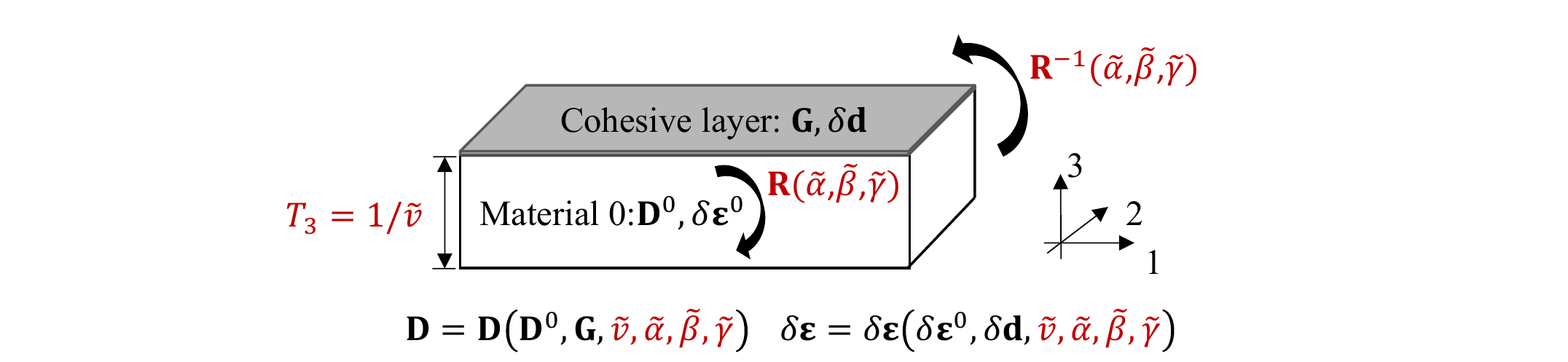}
	\caption{Illustration of the cohesive building block. $\tilde{v}$ is the reciprocal length parameter, $T_3$ is the effective thickness of the building block, and ($\tilde{\alpha},\tilde{\beta},\tilde{\gamma}$) are the rotation angles.}
	\label{fig:buildingblock}
\end{figure}
In a generic cohesive building block shown in Figure \ref{fig:buildingblock}, a non-negative reciprocal length parameter $\tilde{v}$ is introduced for the cohesive layer to represent its size effect. Physically, one can interpret $\tilde{v}$ as the inverse of effective thickness $T_3$ of the undeformed building block in the direction normal to the cohesive layer:
\begin{equation}\label{eq:vthick}
\tilde{v} = \dfrac{1}{T_{3}}.
\end{equation}
Since the deformation of the cohesive layer is characterized by its separation displacement, this reciprocal length parameter $\tilde{v}$ will be further used to transform the displacement vector to the strain tensor contributed to the building block. {It is noted that when $\tilde{v}=0$, the building block has infinite effective thickness and sees no effect from the cohesive layer, indicating that it is perfectly bonded.}

For the $p$-th cohesive layer within the $q$-th cohesive network, its reciprocal length parameter $\tilde{v}_q^p$ is designed to be activated by the rectified linear unit (ReLU) $a(\cdot)$:
\begin{equation}\label{eq:v}
\tilde{v}_q^p = \dfrac{a(\tilde{z}_q^p)}{L} = \dfrac{\max(\tilde{z}_q^p,0)}{L},
\end{equation}
where $L$ is the characteristic length of the microstructure, which is a predefined constant. 

{Theoretically, $L$ does not affect the optimum solution. However, an appropriately selected $L$ will help to keep the activations vary in a similar range as the rotation angles, and thus it eases the convergence of the gradient descent algorithm used for the training. The choice of characteristic length $L$ depends on the morphology of the microstructure. For example, in a unidirectional fiber-reinforced composite, $L$ can be chosen as the average diameter of the fibers. While for another polycrystalline material,  $L$ can be chosen as the average grain size. } 

The activation of cohesive layer $\tilde{z}_q^p$ in Eq. (\ref{eq:v}) is a unitless fitting parameter to be learned. Similar to the activation of node $z$ in the original material network, once $\tilde{z}_q^p$ becomes negative during the training, the corresponding cohesive layer is deactivated, and it will not be activated again due to the vanished gradient of the ReLU function. 

Assume the volume of the $q$-th enriched node is given by $\Omega_q$. Since the cohesive layers have zero volume, all the cohesive building blocks in the $q$-th cohesive network share the same volume $\Omega_q$. Following the definition of $\tilde{v}$ in Eq. (\ref{eq:vthick}, one can define the effective area of the $p$-th cohesive layer $\tilde{S}_q^p$ as
\begin{equation}\label{eq:effectiveA}
\tilde{S}_q^p = \tilde{v}_q^p \: \Omega_q.
\end{equation}

Note that the RVE scaling effect due to the existence of deformable interfaces can be considered by scaling the characteristic length $L$ accordingly. For example, assume an RVE is trained to a DMN with a set of activations $\tilde{z}$ under $L$. If the RVE geometry is enlarged linearly by a scale factor of 2 and the cohesive law stays the same, the new DMN can be obtained by simply updating the characteristic length to $2L$ without changing the trained fitting parameters. According to Eq. (\ref{eq:v}), this is also equivalent to decreasing all the reciprocal length parameters $\tilde{v}$ by a factor of 0.5.

\subsection{Cohesive building block}\label{sec:buildingblock}
As depicted in Figure \ref{fig:buildingblock}, a generic cohesive building block has two inputs: a planar cohesive layer and a bulk material 0. The cohesive layer has zero thickness, and its normal is in direction 3. In general,  its constitutive relation for an arbitrary traction-separation law can be linearized as
\begin{equation}\label{eq:cohe}
\Delta\textbf{d} = {\textbf{G}}\Delta\textbf{t} + \delta\textbf{d},
\end{equation}
where $\Delta\textbf{d}$ is the increment of displacement vector, and $\Delta\textbf{t}$ is the increment of traction vector. {The residual displacement vector $\delta\textbf{d}$ will appear in a nonlinear traction-separation law, including the bilinear one used in this work.}  For a 3-dimensional problem, the traction $\textbf{t}$ is in the unit of force per unit area. It has three components: $t_n$, $t_s$, and $t_t$, which represent the normal and the two shear tractions, respectively. Similarly, the displacements in these directions are denoted by $d_n$, $d_s$, and $d_t$.  Moreover, the compliance matrix $\textbf{G}$ relates the traction vector to the displacement vector under a linear elastic setting, which has 6 independent components in 3D due to symmetry:
\begin{equation}
{\textbf{G}} = \begin{Bmatrix}
{G}_{nn}&{G}_{ns}&{G}_{nt}\\
&{G}_{ss}&{G}_{st}\\
sym&&{G}_{tt}\\
\end{Bmatrix}.
\end{equation}
The stiffness matrix ${\textbf{K}}$ is defined as the inverse of ${\textbf{G}}$:
\begin{equation}
{\textbf{K}}= {\textbf{G}}^{-1}=\begin{Bmatrix}
{K}_{nn}&{K}_{ns}&{K}_{nt}\\
&{K}_{ss}&{K}_{st}\\
sym&&{K}_{tt}\\
\end{Bmatrix}.
\end{equation}

{Note that for history-dependent traction-separation laws, internal variables will also be carried at the cohesive layer, while the incremental constitutive relation $\ref{eq:cohe}$ still applies at each loading step. For layers with perfectly plasticity, the stiffness in certain direction may become zero, and the corresponding elements in the compliance matrix $\textbf{G}$ will be infinite. To avoid the potential numerical issues,  a tiny stiffness number is recommended to be added to the diagonal terms in $\textbf{K}$. The same practice is also adopted in the paper for interfacial failure analysis (see Eq. (\ref{eq:smallK}), when the cohesive layer is fully separated or damaged.}

In a small-strain setting, the Cauchy stress $\boldsymbol{\sigma}$ and the infinitesimal strain $\boldsymbol{\varepsilon}$ are used as stress and strain measures, respectively. Moreover, the stress and strain are vectorized under the Mandel notation:
\begin{equation}
\boldsymbol{\sigma}=\{\utilde{\sigma}_{11},\utilde{\sigma}_{22},\utilde{\sigma}_{33},\sqrt{2}\utilde{\sigma}_{23},\sqrt{2}\utilde{\sigma}_{13},\sqrt{2}\utilde{\sigma}_{12}\}^T=\{\sigma_{1},\sigma_{2},\sigma_{3},\sigma_{4},\sigma_{5},\sigma_{6}\}^T 
\end{equation}
and
\begin{equation*}
\boldsymbol{\varepsilon}=\{\utilde{\varepsilon}_{11},\utilde{\varepsilon}_{22},\utilde{\varepsilon}_{33},\sqrt{2}\utilde{\varepsilon}_{23},\sqrt{2}\utilde{\varepsilon}_{13},\sqrt{2}\utilde{\varepsilon}_{12}\}^T=\{\varepsilon_{1},\varepsilon_{2},\varepsilon_{3},\varepsilon_{4},\varepsilon_{5},\varepsilon_{6}\}^T,
\end{equation*}
where the subscripts 3, 4, and 5 denote the shear directions. Equipped with these conventions, the constitutive relation of Material 0 can be written as
\begin{equation}
\Delta\boldsymbol{\varepsilon}^0 = {\textbf{D}^0}\Delta\boldsymbol{\sigma}^0 + \delta\boldsymbol{\varepsilon}^0,
\end{equation}
where $\Delta\boldsymbol{\varepsilon}^0$ and $\Delta\boldsymbol{\sigma}^0$  are the strain and stress increments, respectively. $\textbf{D}^0$ is its compliance matrix, which is symmetric and has 21 independent entries for a 3-dimensional problem. Additionally, $\delta\boldsymbol{\varepsilon}^0$ is the residual strain.

Three operations are introduced in the building block to propagate the mechanical information including the compliance matrix and residual strain/displacement. First, material 0 is rotated and the obtained constitutive relation is
\begin{equation}
\Delta\boldsymbol{\varepsilon}^{r} = {\textbf{D}}^{r}\Delta\boldsymbol{\sigma}^r + \delta\boldsymbol{\varepsilon}^{r}.
\end{equation} 
Then rotated material 0 is homogenized together with the cohesive layer, and the new constitutive relation is
\begin{equation}
\Delta\bar{\boldsymbol{\varepsilon}} = \bar{\textbf{D}}\Delta\bar{\boldsymbol{\sigma}} + \delta\bar{\boldsymbol{\varepsilon}}.
\end{equation}
To recover the initial orientation of material 0, the homogenized material is rotated back, which gives the overall constitutive relation of the cohesive building block:
\begin{equation}
\Delta\boldsymbol{\varepsilon} = {\textbf{D}}\Delta\boldsymbol{\sigma} + \delta\boldsymbol{\varepsilon}.
\end{equation}
The associated constitutive relations of the inputs, intermediate steps (i, ii), and output step (iii) are summarized in Table \ref{table:conre}.
\begin{table}[!t]
	\captionabove{Constitutive relations in the cohesive building block. The final expressions of the compliance matrix $\textbf{D}$ and the residual strain $\delta \boldsymbol{\varepsilon}$ of are given in Eq. (\ref{eq:blockfinal}) and (\ref{eq:blockfinal1})， respectively.} 
	\centering 
	\label{table:conre} 
	{\tabulinesep=1.0mm
		\begin{tabu}{l  l l} 
			\hline 
			\multirow{2}{*}{Inputs}&Cohesive layer&$\Delta\textbf{d} = {\textbf{G}}\Delta\textbf{t} + \delta\textbf{d}$\\
			&Material 0 & $\Delta\boldsymbol{\varepsilon}^0 = {\textbf{D}^0}\Delta\boldsymbol{\sigma}^0 + \delta\boldsymbol{\varepsilon}^0$\\
			\hline
			\multirow{2}{*}{Intermediate steps}&(i) Rotation of material 0&$\Delta\boldsymbol{\varepsilon}^{r} = {\textbf{D}}^{r}\Delta\boldsymbol{\sigma}^r + \delta\boldsymbol{\varepsilon}^{r}$\\
			&(ii) Homogenization&$\Delta\bar{\boldsymbol{\varepsilon}} = \bar{\textbf{D}}\Delta\bar{\boldsymbol{\sigma}} + \delta\bar{\boldsymbol{\varepsilon}}$\\
			\hline
			Output step &(iii) Rotation of homogenized material&$\Delta\boldsymbol{\varepsilon} = {\textbf{D}}\Delta\boldsymbol{\sigma} + \delta\boldsymbol{\varepsilon}$\\
			\hline
	\end{tabu}}
\end{table}

In the generic building block, the overall input-output function defined on the compliance matrix can be generally written as
\begin{equation}
\textbf{D} = \textbf{D}\left( \textbf{D}^0,\textbf{G},\tilde{z},\tilde{\alpha},\tilde{\beta},\tilde{\gamma}\right)
\end{equation}
and the one for the residual strain/displacement is
\begin{equation}
\delta\boldsymbol{\varepsilon} = \delta\boldsymbol{\varepsilon}\left( \delta\boldsymbol{\varepsilon}^0,\delta\textbf{d} ,\tilde{z},\tilde{\alpha},\tilde{\beta},\tilde{\gamma}\right).
\end{equation}
Analytical solutions of the input-output functions can be derived following the procedure described below.

\text{}

\textbf{{(i) Rotation of  material 0:} }
The rotation operation is defined by the rotation angles ($\tilde{\alpha}, \tilde{\beta}, \tilde{\gamma}$).  The rotation matrix $\textbf{R}$ can be decomposed as a product of three element rotation matrices,
\begin{equation}\label{eq:smallR}
\textbf{R}(\tilde{\alpha}, \tilde{\beta}, \tilde{\gamma}) = \textbf{X}(\tilde{\alpha})\textbf{Y}(\tilde{\beta})\textbf{Z}(\tilde{\gamma}).
\end{equation} 
The expressions of these elementary rotation matrices can be found in \ref{ap:ap0}. The compliance matrix $\textbf{D}^r$ and residual strain $\delta\boldsymbol{\varepsilon}^{r}$ of the rotated material 0 are
\begin{equation}\label{eq:step1}
\textbf{D}^r = \textbf{R}^{-1}(\tilde{\alpha}, \tilde{\beta}, \tilde{\gamma}){\textbf{D}}^0\textbf{R}(\tilde{\alpha}, \tilde{\beta}, \tilde{\gamma}),\quad \delta\boldsymbol{\varepsilon}^{r} = \textbf{R}^{-1}(\tilde{\alpha}, \tilde{\beta}, \tilde{\gamma})\delta\boldsymbol{\varepsilon}^0.
\end{equation}

\text{}

\textbf{(ii) Homogenization:} 
The equilibrium condition at the interface poses that the tractions on the cohesive layer should satisfy
\begin{equation}
\Delta t_{n}=\Delta \bar{\sigma}_3,\quad \sqrt{2}\Delta t_{s}=\Delta \bar{\sigma}_4,\quad \sqrt{2}\Delta t_{t}=\Delta \bar{\sigma}_5.
\end{equation}
For the rotated material 0, the cohesive layer does not apply any displacement or stress constraints on its 1, 2, and 6 directions. Along with the equilibrium condition, we have
\begin{equation}
\Delta\boldsymbol{\sigma}^r = \Delta\bar{\boldsymbol{\sigma}},\quad \Delta\varepsilon^r_1=\Delta\bar{\varepsilon}_1,\quad \Delta\varepsilon^r_2=\Delta\bar{\varepsilon}_2,\quad \Delta\varepsilon^r_6=\Delta\bar{\varepsilon}_6.
\end{equation}
The contribution of displacement vector at the cohesive layer $\Delta\textbf{d}$ to the homogenized strain $\Delta\bar{\boldsymbol{\varepsilon}}$ is scaled by the reciprocal length parameter $\tilde{v}$, which is interpreted as the inverse of effective thickness of the building block in 3 direction. The unknown components of the homogenized strain can be computed as
\begin{equation}
\Delta\bar{\varepsilon}_3 = \Delta\varepsilon^r_3 + \tilde{v}\Delta d_n,\quad \Delta\bar{\varepsilon}_4 = \Delta\varepsilon^r_4 + \dfrac{\tilde{v}\Delta d_s}{\sqrt{2}},\quad \Delta\bar{\varepsilon}_5 = \Delta\varepsilon^r_5 + \dfrac{\tilde{v}\Delta d_t}{\sqrt{2}}.
\end{equation}
Therefore, the compliance matrix and residual strain after the homogenization step are derived to be
\begin{equation}\label{eq:step2}
\bar{\textbf{D}} = \textbf{D}^r + \tilde{v} \tilde{\textbf{G}},\quad \delta\bar{\boldsymbol{\varepsilon}} =  \delta\boldsymbol{\varepsilon}^r+\tilde{v}\delta\tilde{\textbf{d}},
\end{equation}
where $\tilde{\textbf{G}}$ and $\delta\tilde{\textbf{d}}$ are the modified compliance matrix and displacement vector of the cohesive layer:
\begin{equation}\label{eq:resize}
\tilde{\textbf{G}} = \begin{Bmatrix}
0 & & & & & \\
&0& & & & \\
& & {G}_{nn}&{G}_{ns}/\sqrt{2} &{G}_{nt}/\sqrt{2} & \\
& & & {G}_{ss}/2&{G}_{st}/2 & \\
& &sym & & {G}_{tt}/2& \\
& & & & & 0
\end{Bmatrix},\quad
\delta \tilde{\textbf{d}} = \begin{Bmatrix}
0  \\
0\\
\delta d_n\\
\delta d_s/\sqrt{2}\\
\delta d_t/\sqrt{2}\\
0
\end{Bmatrix}.
\end{equation}
Although it is equivalent to derive the homogenization functions based on the stiffness matrices, in practice, the expressions based on compliance matrices are more concise since all the constraints posed by the cohesive layer are acting on the stress components. 

\text{}

\textbf{(iii) Rotation of homogenized material:} 
The orientation of the material 0 is recovered by an inverse rotation $\textbf{R}^{-1}(\tilde{\alpha}, \tilde{\beta}, \tilde{\gamma})$ on the homogenized material, and the overall compliance matrix and residual strain of the cohesive building block are
\begin{equation}\label{eq:step3}
{\textbf{D}} = \textbf{R}(\tilde{\alpha}, \tilde{\beta},\tilde{\gamma})\bar{\textbf{D}}\textbf{R}^{-1}(\tilde{\alpha}, \tilde{\beta}, \tilde{\gamma}),\quad \delta{\boldsymbol{\varepsilon}} = \textbf{R}(\tilde{\alpha}, \tilde{\beta},\tilde{\gamma})\delta\bar{\boldsymbol{\varepsilon}}.
\end{equation}

By combining Eq. \ref{eq:step1}, \ref{eq:step2}, and \ref{eq:step3}, the input-output functions of the cohesive building block are derived to be
\begin{equation}\label{eq:blockfinal}
{\textbf{D}} = {\textbf{D}}\left({\textbf{D}}^0,\textbf{G},\tilde{v},\tilde{\alpha}, \tilde{\beta},\tilde{\gamma}\right)={\textbf{D}}^0+\tilde{v}\textbf{R}(\tilde{\alpha}, \tilde{\beta},\tilde{\gamma})\tilde{\textbf{G}}\textbf{R}^{-1}(\tilde{\alpha}, \tilde{\beta}, \tilde{\gamma})
\end{equation}
and
\begin{equation}\label{eq:blockfinal1}
\delta{\boldsymbol{\varepsilon}} =\delta{\boldsymbol{\varepsilon}}\left(\delta\boldsymbol{\varepsilon}^0,\delta{\textbf{d}},\tilde{v},\tilde{\alpha}, \tilde{\beta},\tilde{\gamma}\right)=\delta\boldsymbol{\varepsilon}^0+\tilde{v}\textbf{R}(\tilde{\alpha}, \tilde{\beta},\tilde{\gamma})\delta\tilde{\textbf{d}}.
\end{equation}
Apparently, if $\tilde{v}=0$, the cohesive building block sees no effect from the cohesive layer.

Analytical solutions under finite-strain formulation also exist, and more details about the derivation are given in \ref{ap:ap1}. However, all examples presented in this paper are based on small-strain formulation.

\subsection{Data transfer between cohesive and material networks}
By applying the input-output functions of the cohesive building block iteratively as illustrated in Figure \ref{fig:framework}, the compliance matrix at the top of the cohesive network for the $q$-th enriched node with $N_c$ cohesive layers is obtained as
\begin{equation}\label{eq:Dcoh}
\textbf{D}_q^{N_c} = \textbf{D}_q^{0} + \sum_{p=1}^{N_c}\tilde{v}_q^p\textbf{R}(\tilde{\alpha}_q^p, \tilde{\beta}_q^p,\tilde{\gamma}_q^p)\tilde{\textbf{G}}_q^p\textbf{R}^{-1}(\tilde{\alpha}_q^p, \tilde{\beta}_q^p, \tilde{\gamma}_q^p),
\end{equation}
and the residual strain is
\begin{equation}\label{eq:decoh}
\delta{\boldsymbol{\varepsilon}}_q^{N_c} = \delta\boldsymbol{\varepsilon}_q^{0} + \sum_{p=1}^{N_c}\tilde{v}_q^p\textbf{R}(\tilde{\alpha}_q^p, \tilde{\beta}_q^p,\tilde{\gamma}_q^p)\delta\tilde{\textbf{d}}_q^p,
\end{equation}
where $\textbf{D}_q^{0} $ and $\delta\boldsymbol{\varepsilon}_q^{0}$ are the compliance matrix and residual strain of the base material of the $q$-th cohesive network, respectively. Meanwhile, $\tilde{\textbf{G}}_q^p$ and $\delta\tilde{\textbf{d}}_q^p$ are modified from the compliance matrix and displacement vector of the $p$-th cohesive layer based on Eq \ref{eq:resize}. 

The information at the top node of the cohesive network is transferred to the corresponding enriched node in the bottom layer ($i=N$) of the original material network. As the stiffness tensor $\textbf{C}$ and residual stress $\delta\boldsymbol{\sigma}$ are primarily used in the material network, the following operations are performed:
\begin{equation}
\textbf{C}_N^{\mathcal{J}(q)} = \left(\textbf{D}_q^{N_c}\right)^{-1}, \quad \delta\boldsymbol{\sigma}_N^{\mathcal{J}(q)}=-\left(\textbf{D}_q^{N_c}\right)^{-1}\delta{\boldsymbol{\varepsilon}}_q^{N_c}.
\end{equation}
The definition of the mapping function $\mathcal{J}(q)$ is provided in Eq. \ref{eq:jq}.

\section{Multi-stage offline training strategy}\label{sec:multistage}
{A two-stage training strategy is proposed in this work. The first stage learns the essential phase topology with no interfacial effect, while the second stage learns the length scale and orientation of the cohesive layers. In comparison to learning all the fitting parameters in one single step, the two-stage straining strategy decouples the interfacial effect from phase interactions, so that it has fewer parameters at each stage. Overall, it makes the landscape of the cost function less complicated for the learning algorithms to converge.}

\begin{figure}[!t]
	\centering
	\graphicspath{{Figures/}}
	\includegraphics[clip=true,trim = 4.2cm 2.8cm 4.0cm 3.3cm, width = 0.9\textwidth]{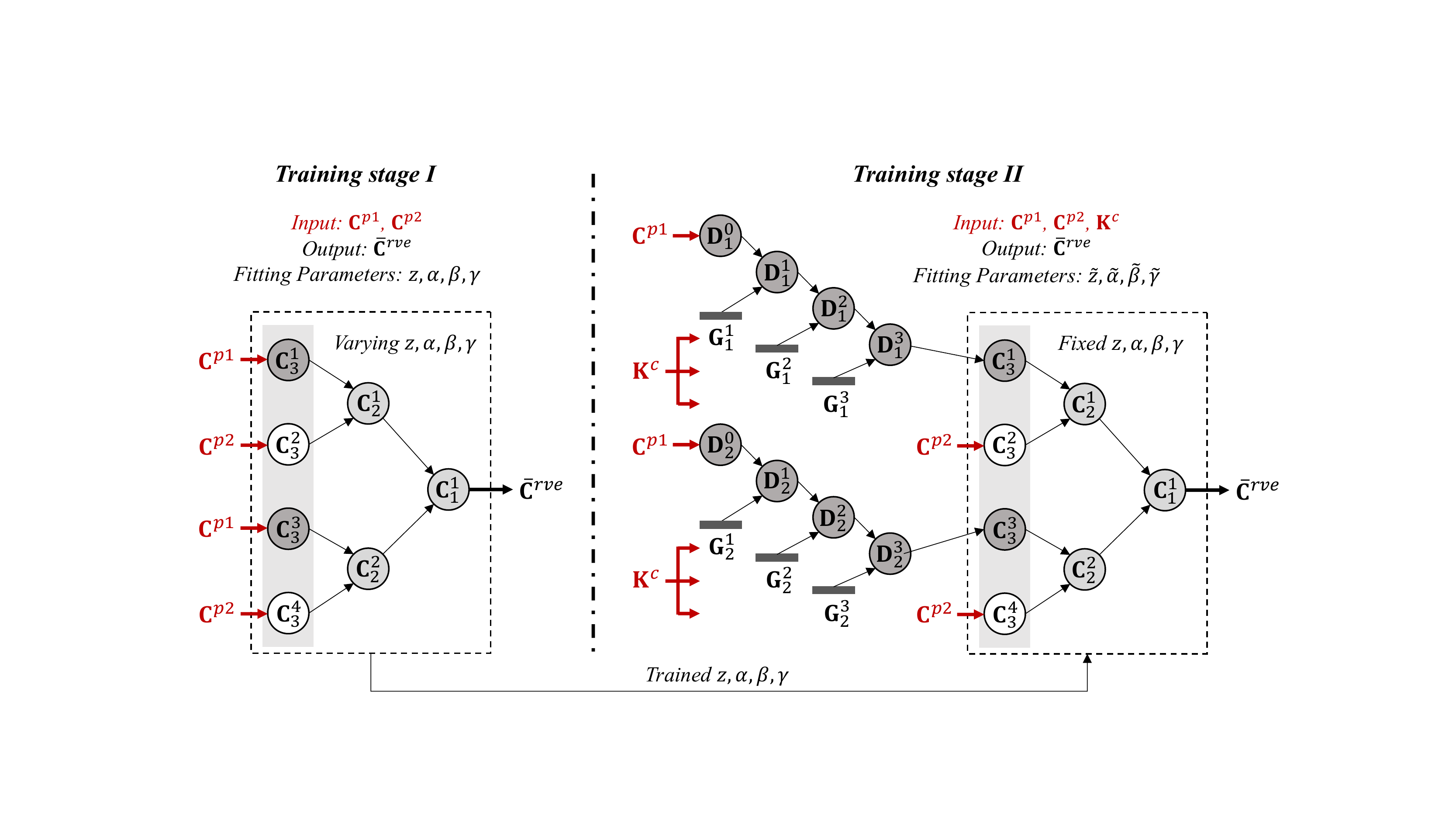}
	\caption{Illustration of the multi-stage training strategy for deep material network with cohesive layers. The number of layers is $N=3$. The number of enriched nodes is $N_e=2$, and the number of cohesive layers in each cohesive network is $N_c=3$. The mapping function is $\mathcal{J}(1)=1,\mathcal{J}(2)=3$. The inputs of training stage I are the stiffness matrices of microscale phases $\textbf{C}^{p1}$ and $\textbf{C}^{p2}$. The extra input of training stage II is the interfacial stiffness matrix $\textbf{K}^c$.}
	\label{fig:mutlistage}
\end{figure}

Figure \ref{fig:mutlistage} illustrated this strategy for a shallow material network with cohesive layers. Its number of layers $N$, number of enriched nodes $N_e$, number of cohesive layers in each cohesive network $N_c$, and mapping function $\mathcal{J}(q)$ between indices of the cohesive network and bottom-layer node in the original material network are listed below:
\begin{equation}
N=3,\quad N_e=2,\quad N_c = 3, \quad \mathcal{J}(1)=1, \quad \mathcal{J}(2)=3.
\end{equation}
Since the offline training processes are based on linear elastic data, the material network only propagates the information of compliance and stiffness matrices.

As shown in the figure, the material network without the enrichment of cohesive networks is first trained for the RVE with perfectly bonded interfaces. The overall stiffness matrix of the multi-layer network  $\bar{\textbf{C}}^{rve}$  for a two-phase RVE can be written as a function of the stiffness matrix from each material phase ($\textbf{C}^{p1}$ and $\textbf{C}^{p2}$) and the fitting parameters ($z, \alpha, \beta, \gamma$):
\begin{equation}\label{eq:final}
\text{Training stage I:}\quad \underbrace{\bar{\textbf{C}}^{rve}}_\text{Output}=\textbf{f}\:( \overbrace{z^{j=1,2,...,2^{N-1}},\alpha_{i=1,2,...,N}^{k=1,2,...,2^{i-1}},\beta_i^k,\gamma_i^k}^\text{Fitting parameters};\: \underbrace{\textbf{C}^{p1},\textbf{C}^{p2}}_\text{Inputs}).
\end{equation}

At training stage II, each active bottom-layer node of phase 1 is linked to a cohesive network to form a new network for the RVE with deformable interfaces. Therefore, the stiffness matrix of the cohesive layers $\textbf{K}^c$ enters as an extra input. Note that the fitting parameters of the two-layer building blocks obtained in the first training stage will be kept constant. With $(\tilde{z},\tilde{\alpha}, \tilde{\beta},\tilde{\gamma})$ as the new fitting parameters, the overall stiffness matrix the multi-layer network  $\tilde{\textbf{C}}^{rve}$ can be written as
\begin{equation}
\text{Training stage II:}\quad
\underbrace{\bar{\textbf{C}}^{rve}}_\text{Output}=\tilde{\textbf{f}}\:( \overbrace{\tilde{z}^{p=1,2,...,N_c}_{q=1,2,...,N_e}, \tilde{\alpha}^p_q, \tilde{\beta}^p_a, \tilde{\gamma}^p_q}^\text{Fitting parameters};\:\underbrace{\textbf{C}^{p1},\textbf{C}^{p2},\textbf{K}^c}_\text{Inputs}, \underbrace{z^j,\alpha_i^k,\beta_i^k,\gamma_i^k)}_\text{Constants}.
\end{equation}
For both stages, the cost function $J$ is defined upon the mean square error (MSE):
\begin{equation}\label{eq:mse}
J =\dfrac{1}{2N_s} \sum_{s=1}^{N_s}\dfrac{||\bar{\textbf{C}}^{dns}_s-\bar{\textbf{C}}^{rve}||^2}{||\bar{\textbf{C}}^{dns}_s||^2},
\end{equation}
where $\bar{\textbf{C}}^{dns}_s$ is the overall stiffness matrix of sample $s$ computed from DNS. $N_s$ is the total number of training samples. The operator $||...||$ denotes the Frobenius matrix norm. While for training stage I, an extra regularization term is added to the cost function to constrain the magnitude of $\{z^j\}$ and makes the optimization problem well-posed \cite{liu2019exploring}. 

Stochastic gradient descent (SGD) is used for minimizing the cost function, while the gradients are calculated through back-propagation algorithm enabled by the analytical solutions derived in Section \ref{sec:buildingblock}. In the rest of this section, details on the sampling and training of each stage are provided.

\subsection{Training stage I}
The inputs $\textbf{C}^{p1}$ and $\textbf{C}^{p2}$ are assigned to the nodes in the bottom layer $N$ in a scheme described as below,
\begin{equation}\label{eq:input}
\textbf{C}_N^{2l-1}=\textbf{C}^{p1},\quad \textbf{C}_N^{2l}=\textbf{C}^{p2}\quad\text{with } l=1,2,...,2^{N-2}.
\end{equation}
This scheme is applicable to two-phase materials. The material network based on binary-tree structure can also handle materials with more than two phases, however, a different assigning scheme needs to be developed.

In the sampling of inputs, both material phases are assumed to be orthotropically elastic, whose compliance matrices can be written as
\begin{equation}
\textbf{D}^{pi}=\left(\textbf{C}^{pi}\right)^{-1}=
\begin{Bmatrix}
1/E_{11}^{pi}&-\nu_{12}^{pi}/E_{22}^{pi}&-\nu_{31}^{pi}/E_{11}^{pi}&&&\\
&1/E_{22}^{pi}&-\nu_{23}^{pi}/E_{33}^{pi}&&&\\
\text{sym}&&1/E_{33}^{pi}&&&\\
&&&1/(2G_{23}^{pi})&&\\
&&&&1/(2G_{31}^{pi})&\\
&&&&&1/(2G_{12}^{pi})\\
\end{Bmatrix},
\end{equation}
where $pi$ is $p1$ or  $p2$. To initiate material anisotropy, the tension moduli of each phase are first randomly sampled:
\begin{equation}
\log_{10}(E_{11}^{pi}),\enskip \log_{10}(E_{22}^{pi}),\enskip \log_{10}(E_{33}^{pi}) \sim U[-1, 1],
\end{equation}
where $U$ represents the uniform distribution. The moduli of phase 2 will be rescaled to create phase contrasts, and the rescaling factor $\bar{E}^{p2}$ is sampled from
\begin{equation}
\log_{10}(\bar{E}^{p2})\sim U[-3, 3].
\end{equation}
With the rescaling factor, the tension moduli of each phase are updated,
\begin{equation}
E_{kk}^{p2} \leftarrow \dfrac{\bar{E}^{p2}}{(E_{11}^{p2}E_{22}^{p2}E_{33}^{p2})^{1/3}}E_{kk}^{p2}\quad \text{with } k=1,2,3.
\end{equation}
After all the tension moduli are determined, the shear moduli are sampled from
\begin{equation}
\dfrac{G_{12}^{pi}}{\sqrt{E_{11}^{pi}E_{22}^{pi}}},\enskip \dfrac{G_{23}^{pi}}{\sqrt{E_{22}^{pi}E_{33}^{pi}}},\enskip \dfrac{G_{31}^{pi}}{\sqrt{E_{33}^{pi}E_{11}^{pi}}} \sim U[0.25, 0.5].
\end{equation}
To guarantee the compliance matrices are positive definite, the Poisson's ratios are sampled from
\begin{equation*}
\dfrac{\nu_{12}^{pi}}{\sqrt{E_{22}^{pi}/E_{11}^{pi}}}\sim U(0.0,0.5),\enskip \dfrac{\nu_{23}^{pi}}{\sqrt{E_{33}^{pi}/E_{22}^{pi}}}\sim U(0.0,0.5),\enskip
\dfrac{\nu_{31}^{pi}}{\sqrt{E_{11}^{pi}/E_{33}^{pi}}}\sim U(0.0,0.5).
\end{equation*}
Monte Carlo method is used to sample the input space, which has in total 13 independent random variables. 

At the start of training, the fitting parameters ($z, \alpha, \beta, \gamma$) are initialized randomly following uniform distributions. During the training process, the network can be compressed through two operations: 1) Deletion of the parent node with only one child node; 2) subtree merging based on the similarity search \cite{liu2019exploring}.

\subsection{Training stage II}
In training stage II, the active nodes of phase 1 in the bottom layer are enriched by the cohesive network, while the input $\textbf{C}^{p2}$ is still assigned to the other active nodes of phase 2:
\begin{equation}
\textbf{C}_N^{2l}=\textbf{C}^{p2}\quad\text{with } l=1,2,...,2^{N-2}.
\end{equation}
Meanwhile, the input $\textbf{C}^{p1}$ is assigned to the base material of each cohesive network, and all the cohesive layers receive the same stiffness matrix $\textbf{K}^c$:
\begin{equation}
\textbf{D}_q^0 = \left(\textbf{C}^{p1}\right)^{-1},\quad \textbf{G}_q^p = \left(\textbf{K}^{c}\right)^{-1}\quad\text{with }
q = 1,2,...,N_e;\:p = 1,2,...,N_c.
\end{equation}

The samples generated from training stage I are reused and augmented with new elastic cohesive properties. The cohesive layer is assumed to be isotropic in the two shear directions. In addition, the off-diagonal terms in $\textbf{K}^c$ are set to zero. As a result, $\textbf{K}^c$ has only two independent components $K_{nn}^c$ and $K_{ss}^c$:
\begin{equation}
\textbf{K}^c=\left(\textbf{G}^c\right)^{-1}=\begin{Bmatrix}
	K_{nn}^c&0&0\\
	0&K_{ss}^c&0\\
	0&0&K_{ss}^c\\
\end{Bmatrix}.
\end{equation}
The elastic constant in the normal direction $K_{nn}^c$ is sampled as 
\begin{equation}
\log_{10}(K_{nn}^c\times L) \in U[-3,3],
\end{equation}
where $L$ is the characteristic length of the microstructure, and a discussion about the choice of $L$ can be found in Section \ref{sec:activation}. For the elastic constant in the shear direction $K_{ss}^c$, we have
\begin{equation}
\log_{10}(\dfrac{K_{ss}^c}{K_{nn}^c}) \in U[-1,1].
\end{equation}
Although the new input space has 15 independent variables, only two of them need to be sampled at this stage and are further appended to the previous space. Similarly, Monte Carlo method is used for the sampling.

The fitting parameters $(\tilde{z},\tilde{\alpha}, \tilde{\beta},\tilde{\gamma})$ are initialized randomly, while the parameters ($z, \alpha, \beta, \gamma$) trained from stage I are fixed during stage II, as shown in Figure \ref{fig:mutlistage}.

Model compression can be introduced to reduce the redundancy of the cohesive layers. By looking at Eq. (\ref{eq:Dcoh}) and (\ref{eq:decoh}), one can see that the terms contributed by each cohesive layer are linearly added to the overall compliance matrix or residual strain. This nice property enables the merge of two cohesive layers regardless of their order in the network.  

In each cohesive network, the similarity between any pair of cohesive layers will be evaluated. Let us assume that their fitting parameters are given by ($\tilde{z}^{p},\tilde{\alpha}^{p}, \tilde{\beta}^{p}, \tilde{\gamma}^{p}$) and ($\tilde{z}^{p'},\tilde{\alpha}^{p'}, \tilde{\beta}^{p'}, \tilde{\gamma}^{p'}$). Since the cohesive layers are set to behave isotropically in the shear directions, they are said to be similar if they share the same normal direction. Mathematically, it requires that the absolute value of the 33 component in a matrix $\textbf{R}^{p\text{-}p'}$ is close to 1:
\begin{equation*}
|{R}^{p\text{-}p'}_{33}| \approx 1,
\end{equation*}
with $\textbf{R}^{p\text{-}p'}$ defined as
\begin{equation}
\textbf{R}^{p\text{-}p'} = \textbf{R}(\tilde{\alpha}^p, \tilde{\beta}^p, \tilde{\gamma}^p)\textbf{R}^{-1}(\tilde{\alpha}^{p'}, \tilde{\beta}^{p'}, \tilde{\gamma}^{p'}).
\end{equation}
In this case, the two cohesive layers will be merged, and their new activations are
\begin{equation}
\tilde{z}^{p(new)} = \tilde{z}^p + \tilde{z}^{p'}, \quad \tilde{z}^{p'(new)} = 0.
\end{equation}
The $p'$-th cohesive layer is deactivated, so that the size of the machine-learning model is reduced.

\section{Cohesive law for online interfacial failure analysis}\label{sec:coh}
In the online stage, the DMN with cohesive layers, trained based on linear elastic data, can be extrapolated to consider nonlinear interfacial behavior, such as the failure at the material interface. Specifically, the irreversible cohesive law proposed by Camacho and Ortiz \cite{camacho1996computational,ortiz1999finite} is adopted in this work. Simplification first arises from the assumption that the cohesive surface is isotropic. This indicates that the resistance to sliding is independent of the direction of sliding. and only depends on the magnitude $d_S$:
\begin{equation}
d_S = \sqrt{{d_s}^2+{d_t}^2}.
\end{equation}
The total opening displacement vector $\textbf{d}$ and traction $\textbf{t}$ take the following forms,
\begin{equation}
\textbf{d} = d_n \textbf{n} + \textbf{d}_S, \quad \textbf{t} = t_n \textbf{n} + \textbf{t}_S,
\end{equation}
where $\textbf{n}$ is the unit normal to the cohesive surface. An effective opening displacement $d_m$ is further introduced to simplify the mixed-mode cohesive law, and the free energy density per unit undeformed area $\phi$ is a function of $d_m$,
\begin{equation}
\phi = \phi\left(d_m,\textbf{q}\right),
\end{equation}
where $\textbf{q}$ are some internal variables which describe the irreversible processes. Accordingly, the effective traction is defined as
\begin{equation}
t_m = \dfrac{\partial \phi}{\partial d_m}(d_m,\textbf{q}).
\end{equation}

The compressive stress state should not cause the cohesive layer to fail, so that it does not contribute to the free energy density. In addition, no friction effect is included in the cohesive model. The tensile and compressive cases are considered separately:
\begin{enumerate}
	\item For the tensile case $d_n\geq0$, the effective opening displacement $d_m$ is
	\begin{equation}
	d_m = \sqrt{{d_n}^2+\beta^{2}{d_S}^2},
	\end{equation}
	where the positive parameter $\beta$ defines the ratio of effects from normal and shear displacements. The cohesive law becomes
	\begin{equation}
	\textbf{t} = \dfrac{\partial\phi}{\partial \textbf{d}} = \dfrac{t_m}{d_m}\left(d_n\textbf{n}+\beta^{2}\textbf{d}_S\right).
	\end{equation}
	\item For the compressive case $d_n<0$, the effective opening displacement $d_m$ is
	\begin{equation}
	d_m = \beta|{d_S}|.
	\end{equation}
	The cohesive law can be written as
	\begin{equation}
	\textbf{t} = t_n\textbf{n}+\dfrac{\partial\phi}{\partial \textbf{d}_S} =t_n\textbf{n}+ \dfrac{t_m}{d_m}\beta^{2}\textbf{d}_S.
	\end{equation}
\end{enumerate}

\begin{figure}[!t]
	\centering
	\graphicspath{{Figures/}}
	\includegraphics[clip=true,trim = 6cm 6cm 6cm 7cm, width = 0.9\textwidth]{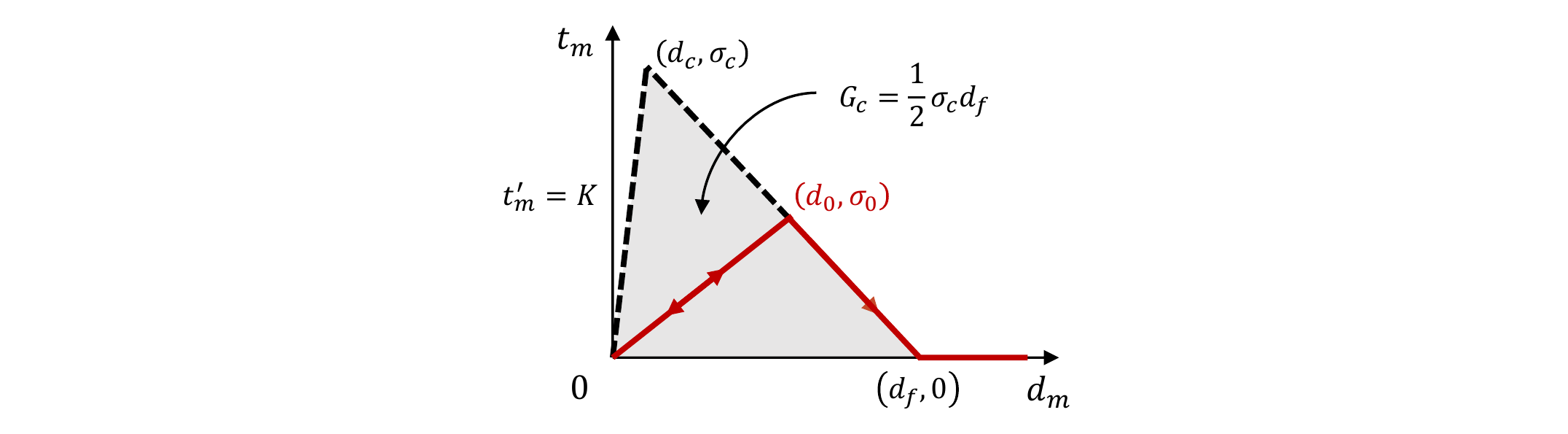}
	\caption{The bilinear cohesive law expressed in terms of the effective opening displacement $d_m$ and the effective traction $t_m$. The initial stiffness is $K$, and the critical energy releasing rate is denoted by $G_c$.}
	\label{fig:cohesive}
\end{figure}
As shown in Figure \ref{fig:cohesive}, a bilinear cohesive law is adopted in this work. In the elastic loading regime, the modulus of cohesive layer is $K$. In the application to be shown in Section \ref{sec:application}, this interfacial modulus is chosen to be a large value compared to the moduli of other material phases, thus, the interface is nearly perfectly bonded at the beginning. The effective opening displacement before the softening regime is denoted by $d_c$, and the corresponding effective traction is $\sigma_c$. The opening displacement at full failure $t_m=0$ is $d_f$.  The critical energy releasing rate $G_c$ is equal to the area under the curve of $t_m$-${d_m}$,
\begin{equation}\label{eq:Gc}
G_c = \dfrac{1}{2}\sigma_cd_f.
\end{equation}

The only internal variable is the effective opening displacement at maximum traction $d_0$. Initially, $d_0$ is equal to $d_c$, 
\begin{equation}
d_0 = d_c,
\end{equation}
and it should decrease along the softening path when $d_m$ exceeds the current $d_0$. Mathematically, the effective traction $t_m$ of the bilinear cohesive law can be calculated by
\begin{equation}\label{eq:effcoh}
t_m =  \dfrac{d_m-d_f}{d_c-d_f}\sigma_c\quad\text{if } d_m = d_0 \text{ and } \dot{d}_m\geq0,
\end{equation}
and 
\begin{equation}
t_m = 
\dfrac{d_m}{d_0}\sigma_0\quad\text{if } d_m < d_0 \text{ or } \dot{d}_m<0.
\end{equation}
In practice, to avoid singularity at full separation of cohesive layer, a linear term $\kappa d_m$ is added to the effective traction, 
\begin{equation}\label{eq:smallK}
t_m \leftarrow t_m + \kappa d_m
\end{equation}
where $\kappa$ is a tiny stiffness parameter chosen as $\kappa/K=1\times10^{-6}$ in this work.

When reloading, the curve will simply trace the unloading path and rejoin the cohesive law at $d_m=d_0$.  Meanwhile, an elastic response is assumed for the compressive case, with the same modulus as the one in the bilinear law,
\begin{equation}
t_n = Kd_n\quad \text{if } d_n < 0.
\end{equation}

{More details on the derivation of the tangent stiffness matrix and viscous regularization for implicit analysis in the online stage are provided in \ref{ap:ap2}.}

\section{Application to unidirectional fiber-reinforced composite} \label{sec:application}
\subsection{Problem settings}
\begin{figure}[!t]
	\centering
	\graphicspath{{Figures/}}
	\subfigure[Geometry and mesh.]{\includegraphics[clip=true,trim = 7.6cm 3.5cm 7.6cm 4.0cm,width=0.44\textwidth]{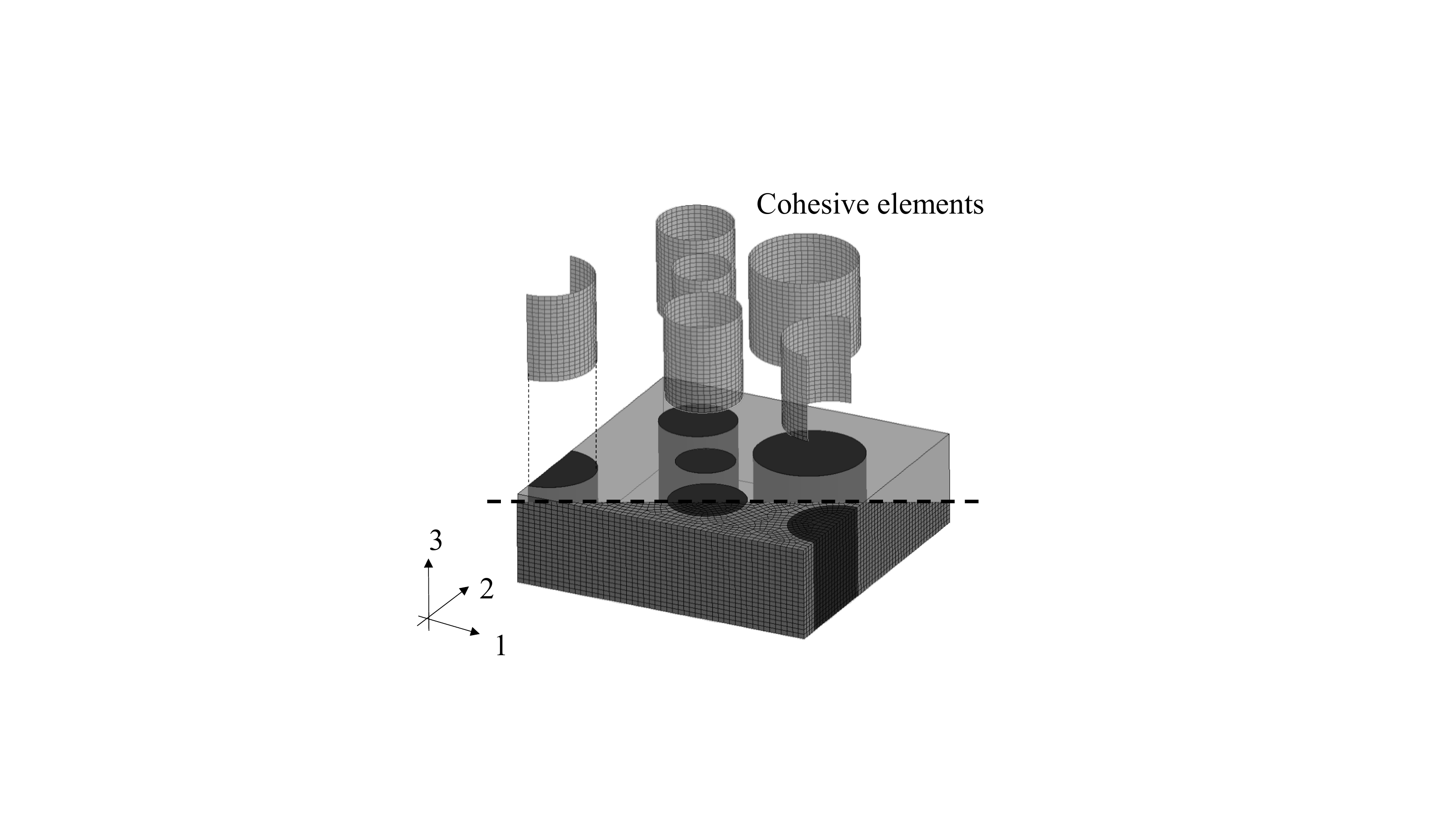}}
	\subfigure[Loading path.]{\includegraphics[clip=true,trim = 9.0cm 4.0cm 9.0cm 4.5cm,width=0.44\textwidth]{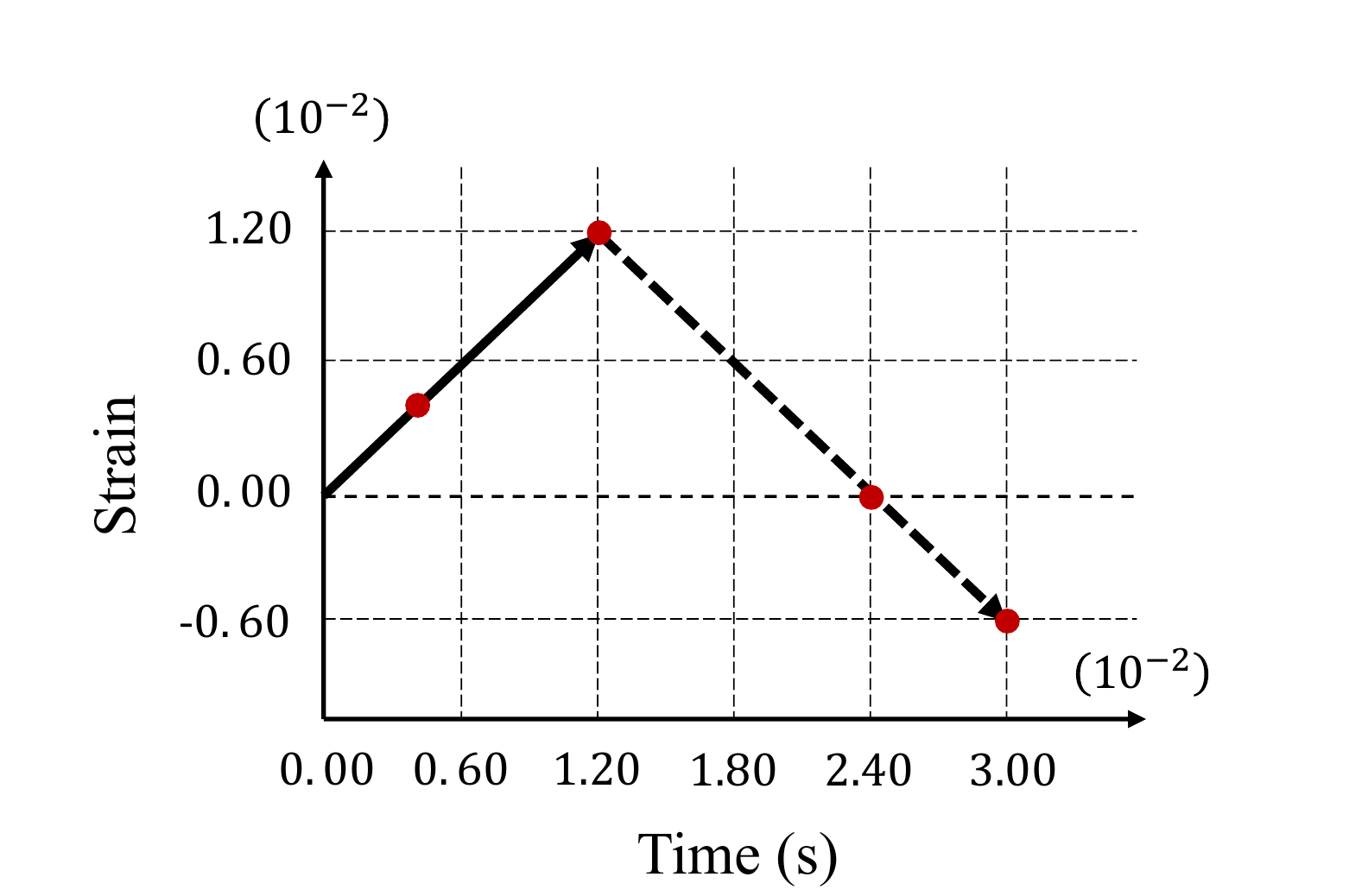}}
	\caption{Unidirectional fiber-reinforced RVE: (a) Volume fraction of the fiber phase is 29.4\%, the FE model has 69048 nodes, 63720 8-node hexahedron elements, and 4320 8-node zero-thickness cohesive elements; (b) History of strain component in the online stage includes both loading and unloading paths. Snapshots of deformed RVEs are taken at t=0.04, 0.12, 0.24, 0.30 s.}
	\label{fig:ud}
\end{figure}

Here the deep material network with cohesive layers is applied to describe a two-phase unidirectional (UD) fiber-reinforced composite. The method is general, and another example of the particle-reinforced composite is briefly discussed in \ref{ap:ap3}. The geometry of the UD RVE is shown in Figure \ref{fig:ud} (a). The volume fraction of the fiber phase is 29.4\%. The average diameter of the fibers is around 2.5 mm, so that the characteristic length $L$, a hyper-parameter for the cohesive networks, is set to be
\begin{equation}
L = 2.5 \text{ mm}.
\end{equation}

In the FE mesh for DNS, there are 69048 nodes, 63720 8-node hexahedron solid elements, and 4320 8-node zero-thickness cohesive elements. All the material parameters used in the online stage are given in Table \ref{table:materialpara}.
\begin{table}[t!]
	\captionabove{Material parameters of the fiber phase, the matrix phase and the interfaces used in the online stage. } 
	\centering 
	\label{table:materialpara} 
	{\tabulinesep=1.0mm
		\begin{tabu}{c c c c c c c} 
			\hline
			\multirow{2}{*}{Fiber}&$E_1$ (GPa) & $\nu_{1}$ & &&\\
			&500 & 0.3 &&& \\ 
			\hline
			\multirow{2}{*}{Matrix}&$E_2$ (GPa) & $\nu_2$ & $\sigma^y_2$ (GPa) &Hardening&\\
			&100 & 0.3 & 0.1 & Eq. (\ref{eq:piecewise_hardening}) &\\
			\hline
			\multirow{2}{*}{Interface}&$K$ (GPa/m) & $\sigma_c$ (GPa) & $G_c$ (GPa$\cdot$m) &$\beta$&$\zeta$ (GPa$\cdot$s)\\
			&$1\times10^{7}$& 0.5 & $2.5\times10^{-6}$ & 0.5 & $2\times10^{-5}$ \\  
			\hline
	\end{tabu}}
\end{table}

The fiber phase is assumed to be linear elastic, and its Young's modulus $E_1$ and Poisson's ratio $\nu_1$ are 500 GPa and 0.3, respectively. Two cases are considered for the matrix phase. In the first case, the matrix phase is linear elastic, with $E_2=100$ GPa and $\nu_2=0.3$. In the second case, the matrix phase is an elasto-plastic material, which obeys von Mises plasticity with piecewise linear hardening. The material yields at 0.1 GPa, and the yield stress $\sigma^Y_{2}$ is a function of the equivalent plastic strain $\varepsilon^p_2$:
\begin{equation}\label{eq:piecewise_hardening}
\sigma^Y_{2}(\varepsilon^p_2)=
\begin{cases}
0.1 + 50{\varepsilon}^{p}_{2} & \varepsilon^p_1 \in [0,0.01) \\
0.4 + 20{\varepsilon}^{p}_{2} & \varepsilon^p_1 \in [0.01,\infty)
\end{cases}
\text{  GPa}.
\end{equation} 

The equivalent modulus ($K\times L$=25000 GPa) of the interface is much larger than the moduli of fiber and matrix phases, indicating that the interface is nearly perfectly bonded at the beginning. Following Eq. (\ref{eq:Gc}), the effective opening displacement at complete failure $d_f$ can be computed from the critical effective traction $\sigma_c$ and the critical energy releasing rate $G_c$:
\begin{equation}
d_f = \dfrac{2G_c}{\sigma_c}=0.01\text{ mm}.
\end{equation}
To restrain the rate-dependency from the numerical stabilization in Eq. \ref{eq:vis}, the viscosity-like parameter $\zeta$ (see the definition in Eq. (\ref{eq:vis})) is set to a small value.

For both training stages, there are 400 training samples and 100 test samples. {For each sample, a set of linear elastic DNS of the UD RVE under 6 orthogonal loading conditions is performed, which typically takes 182 s using LS-DYNA\textregistered \text{} RVE package on 10 Intel\textregistered $ $ Xeon\textregistered $ $ CPU E5-2640 v4 2.40 GHz processors \footnote{The same machine is used for all other computations in this work.}. With 1000 samples in total, the offline data generation process costs 50 h. This time can be further reduced by distributing the jobs to more processors.}

In the SGD algorithm, the mini-batch size is chosen as 20, so that there are 20 learning steps in each epoch. Additionally, a Bold driver is applied to dynamically adjust the learning rate during the training. {The DMN framework integrated offline sampling, definition of the DMN data structure, offline training (e.g. backpropagation), and online prediction. Currently, it is  implemented in Python, with necessary libraries for matrix manipulation and linear algebra.}

\subsection{Offline training results}
In training stage I, the original material networks with three different depths, $N=5,7,$ and $9$, are considered. For $N=5$ and 7, the networks were trained for 20000 epochs. Since the network with $N=9$ has more fitting parameters, it was trained for 40000 epochs. The training and test errors are listed in Table \ref{table:cfrp-ud}, where the error of sample $s$ is defined as
\begin{equation}
e_s=\dfrac{||\bar{\textbf{C}}^{dns}_s-\bar{\textbf{C}}^{rve}_s||}{||\bar{\textbf{C}}^{dns}_s||}.
\end{equation}

\begin{figure} [!t]
	\centering
	\graphicspath{{Figures/}}
	\subfigure[$N=5, N_a =4$]{\includegraphics[clip=true,trim = 0.0cm 0.0cm 1.0cm 0.5cm,width=0.25\textwidth]{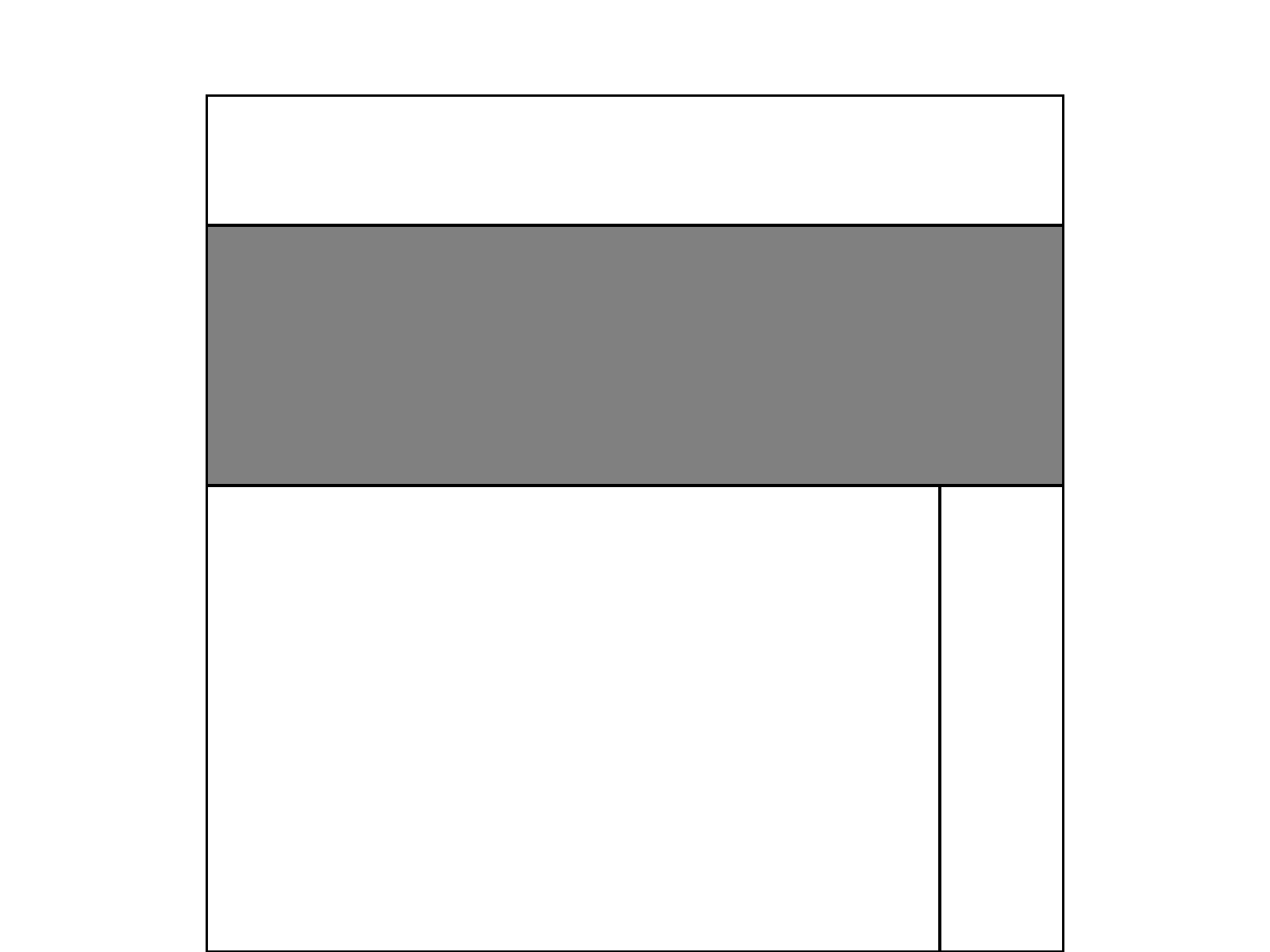}}
	\subfigure[$N=7, N_a =14$]{\includegraphics[clip=true,trim = 0.0cm 0.0cm 1.0cm 0.5cm,width=0.25\textwidth]{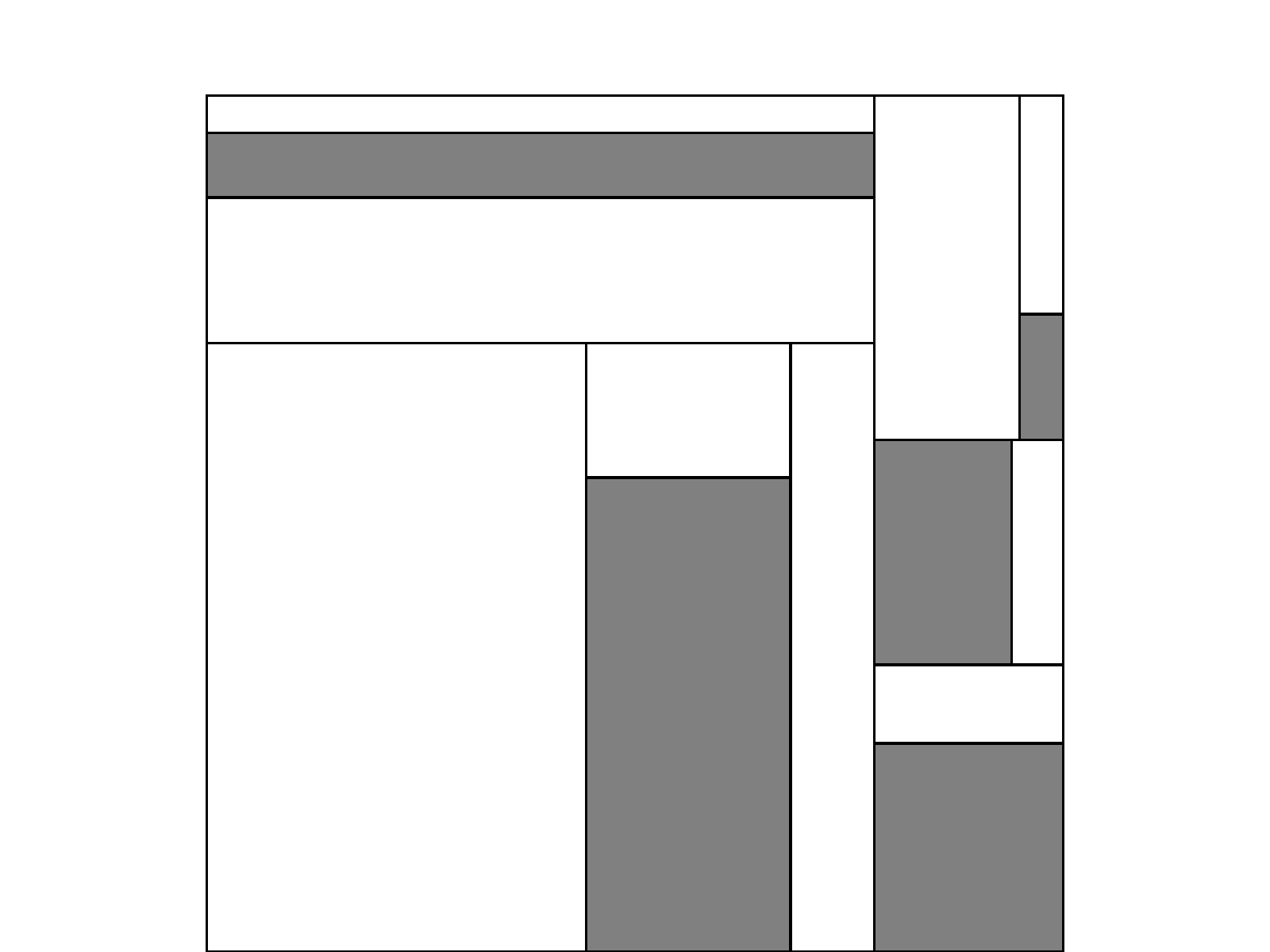}}
	\subfigure[$N=9, N_a =60$]{\includegraphics[clip=true,trim = 0.0cm 0.0cm 1.0cm 0.5cm,width=0.25\textwidth]{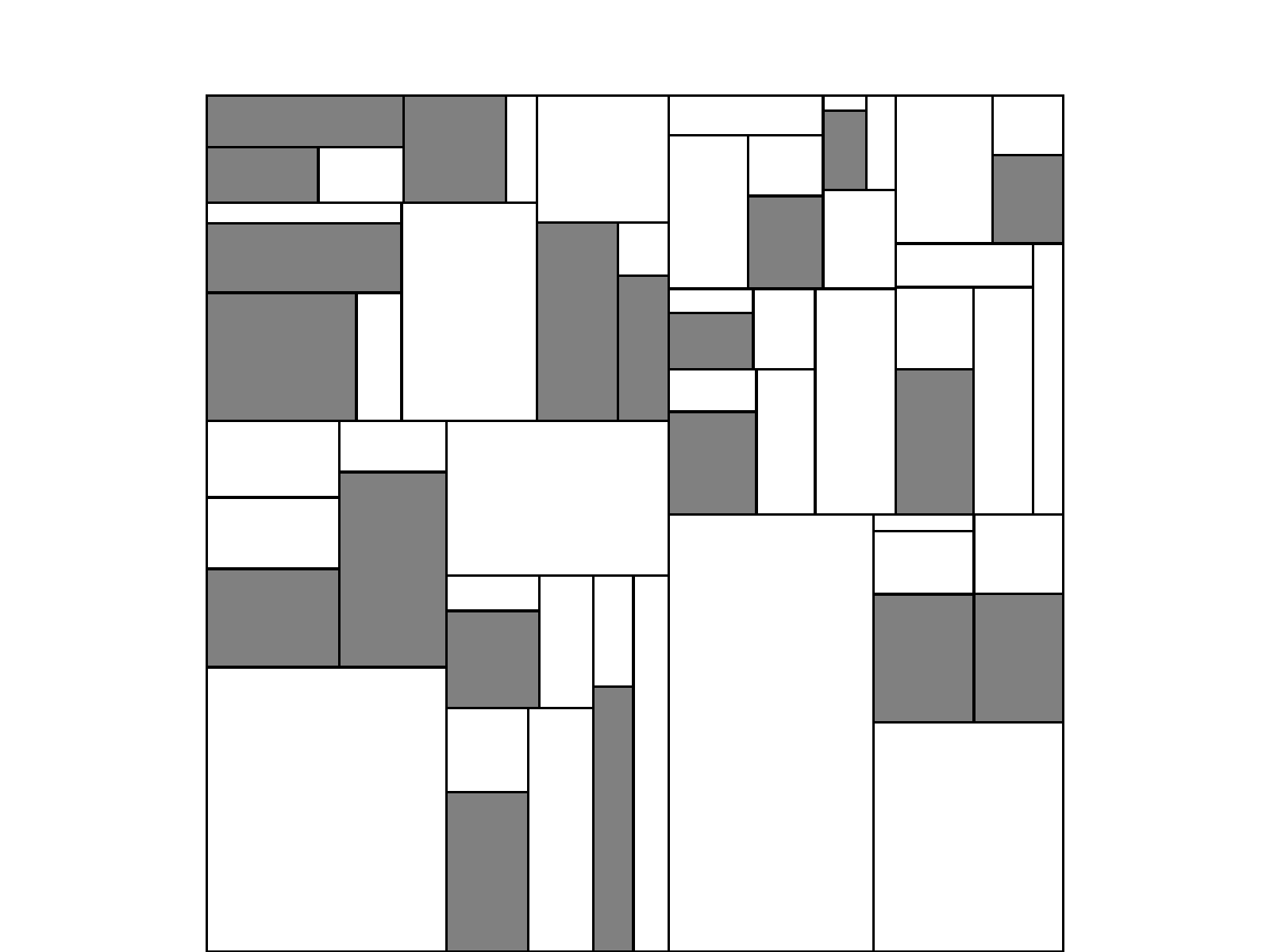}}
	\caption{Treemaps of trained material networks from training stage I. The number of active nodes in the bottom layer $N_a$ is also provided. The dark blocks belong to the fiber phase.}
	\label{fig:udTree}
\end{figure}

To show the fractions of active nodes in the bottom layer, treemap plots of all three DMNs are also provided in Figure \ref{fig:udTree}. For the network with $N=5$, there remains only one active node in the bottom layer for the fiber phase. Evidently, this is not sufficient for capturing the anisotropic behaviors of the UD RVE, and the average test error $\bar{e}^{te}$ is more than 10\%. {One important feature of DMN is that the average training error and the average test error of unseen samples are almost the same as shown in Table \ref{table:cfrp-ud}. With a small amount of data, it requires no regularization scheme (e.g. L2 penalty or Dropout) to avoid the overfitting issue, mostly thanks to the embedded physics in the building block.} More detailed results of training stage I, including the error histories, can be found in \cite{liu2019exploring}.
\begin{table}[!t]
	\captionabove{Training results of UD RVE at stages I and II. Average training error $\bar{e}^{tr}$, average test error $\bar{e}^{tr}$ and maximum test error are provided for each DMN. In stage II, the number of cohesive layers in each cohesive network is $N_c=4$.} 
	\centering 
	\label{table:cfrp-ud} 
	{\tabulinesep=1.0mm
		\begin{tabu}{c | c c c |c c c} 
			\hline 
			&&\textbf{Stage I}&&&\textbf{Stage II}&\\
			&Training $\bar{e}^{tr}$ & Test $\bar{e}^{te}$ & Maximum $e_s^{te}$ &Training $\bar{e}^{tr}$ & Test $\bar{e}^{te}$ & Maximum $e_s^{te}$\\
			\hline
			$N=5$&11.1\%&10.5\%&47.2\%&11.9\%&11.3\%&47.5\%\\
			$N=7$&0.79\%&0.76\%&3.46\%&0.88\%&0.91\%&3.82\%\\
			$N=9$&0.23\%&0.25\%&1.34\%&0.31\%&0.33\%&2.45\% \\
			\hline
	\end{tabu}}
\end{table}
\begin{figure} [!t]
	\centering
	\graphicspath{{Figures/}}
	\subfigure[Different $N$, same $N_c=4$.]{\includegraphics[clip=true,trim = 0.0cm 0.0cm 1.0cm 0.5cm,width=0.44\textwidth]{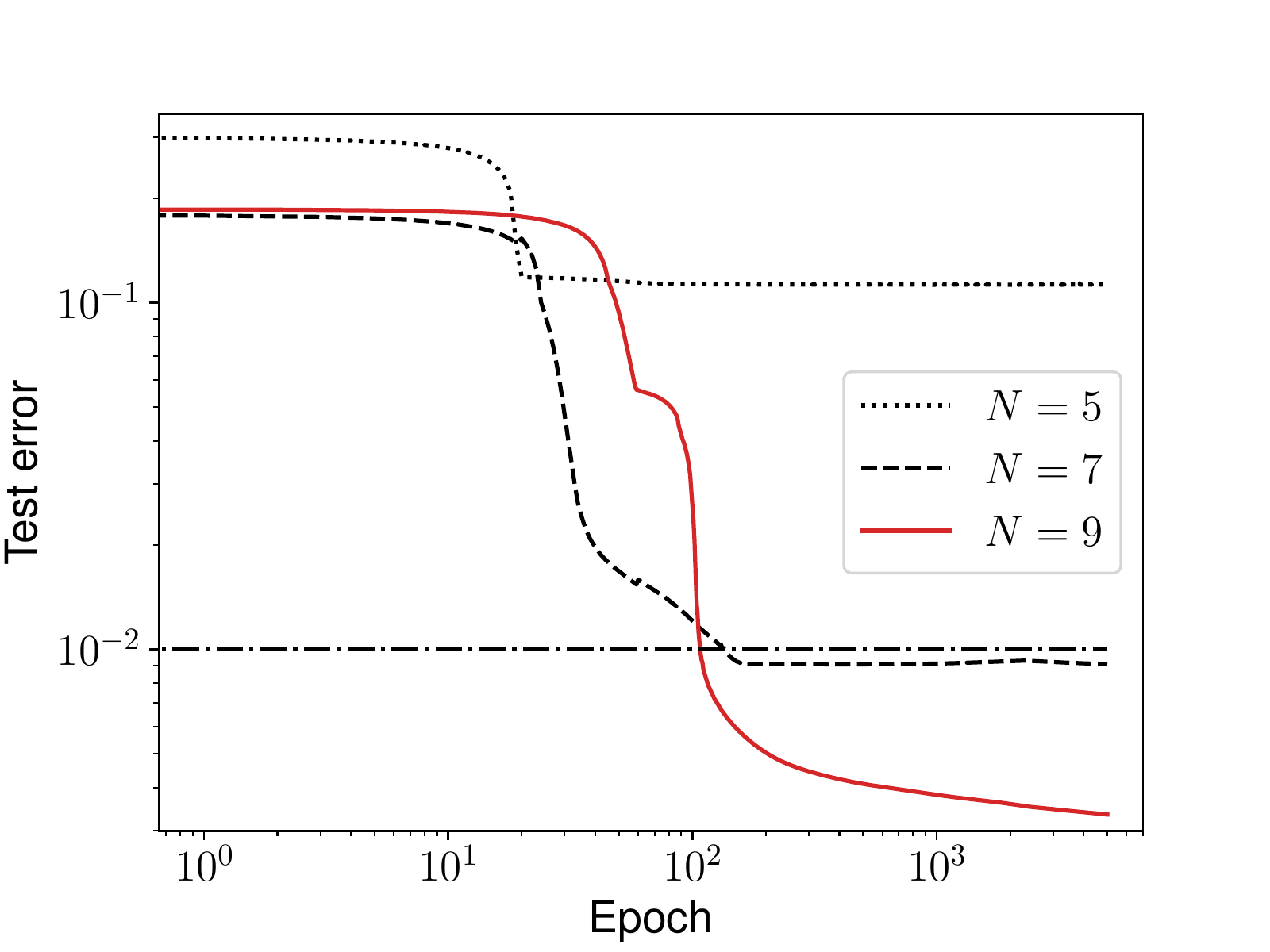}}
	\subfigure[Different $N_c$, same $N=9$.]{\includegraphics[clip=true,trim = 0.0cm 0.0cm 1.0cm 0.5cm,width=0.44\textwidth]{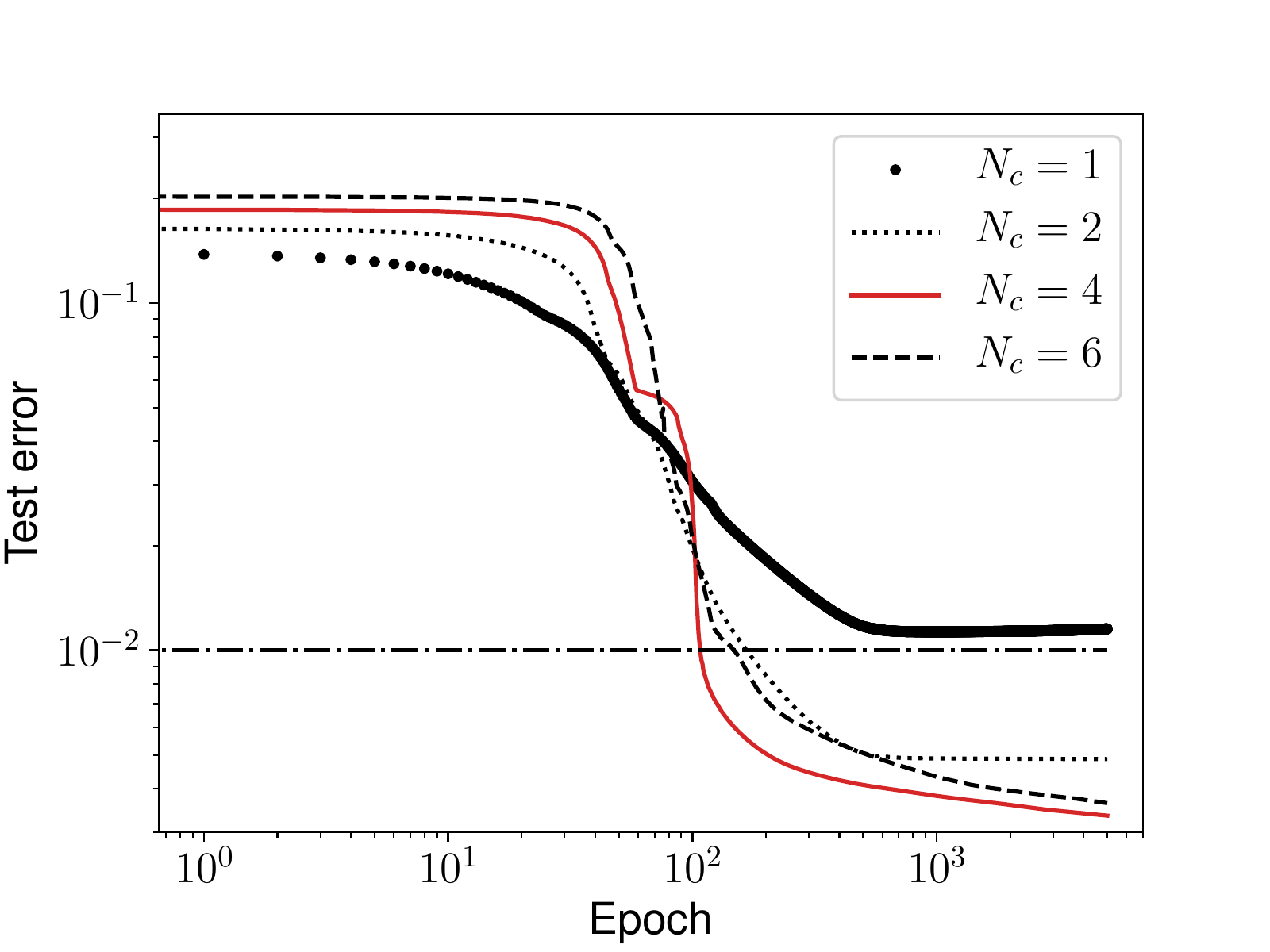}}
	\caption{Histories of test error in training stage II. All the networks are trained for 5000 epochs.}
	\label{fig:history}
\end{figure}

In training stage II, the trained material networks from stage I are enriched with cohesive networks, where the new fitting parameters are also optimized using the SGD algorithm. For all the cases, the networks were trained for 5000 epochs. Figure \ref{fig:history} (a) shows the histories of test error for $N=5$, 7, and 9, with the same number of cohesive layers in each cohesive network $N_c=4$. Their training and test errors at the end of learning are also provided in Table \ref{table:cfrp-ud}. Apparently, the network with $N=9$ offers the most accurate results.

The effect of the number of cohesive layers at each node $N_c$ on the training results has also been investigated for depth $N=9$. We can see from Figure \ref{fig:history} (b) that one cohesive layer $N_c=1$ is not able to capture the interfacial effect accurately. While for $N_c=2$, 4, and 6, the test errors are all reduced to be less than 0.5\%. After 5000 epochs, the network with $N_c=4$ reaches the lowest test error, instead of the one with $N_c=6$. This can be explained by the fact that the network with $N_c=6$ has more redundant cohesive layers, and the hyper-parameter $L=25$ mm may not be optimal anymore. 

It is also interesting to compare the errors from training stage I and II in Table \ref{table:cfrp-ud}. The error at stage I, in any measure, is always less than the one in stage II. This is physically sound as the UD RVE with deformable fiber-matrix interfaces is a more complex material system. Moreover, train stage II inherits the parameters ($z,\alpha,\beta,\gamma$) from stage I, and they are fixed in the training. Following this strategy, the accuracy of enriched network is somehow limited by the original material network.

\begin{figure} [!t]
	\centering
	\graphicspath{{Figures/}}
    \includegraphics[clip=true,trim = 0.0cm 0.0cm 1.0cm 0.5cm,width=0.44\textwidth]{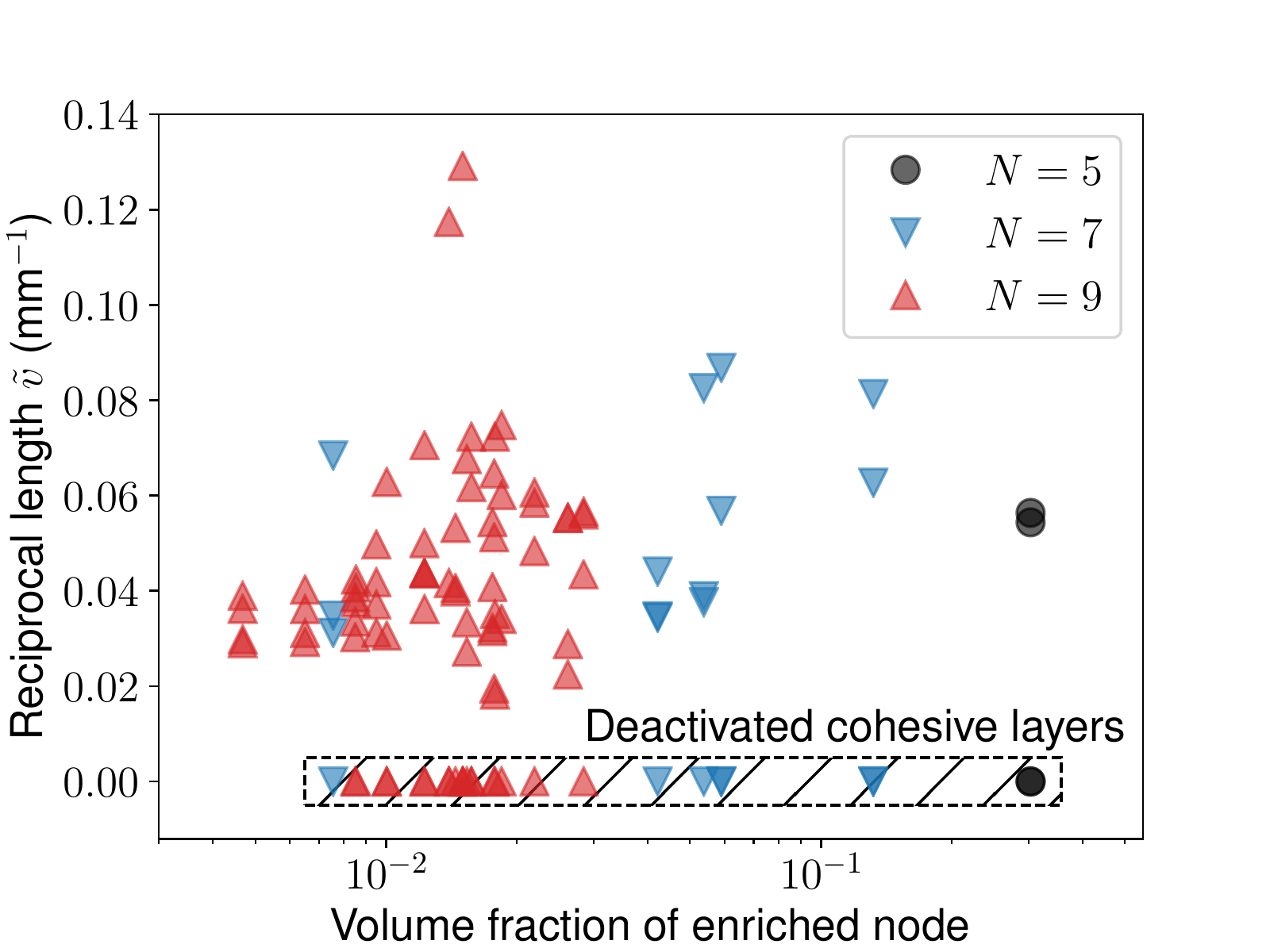}
	\caption{Scatter plots of the cohesive layers for $N=5$, 7, 9 and $N_c=4$ after 5000 epochs of training. The y-axis shows the reciprocal length parameter $\tilde{v}$ of a cohesive layer, and the x-axis shows the volume fraction of the corresponding bottom-layer node enriched by the cohesive layer. The deactivated cohesive layers have $\tilde{v}=0$.}
	\label{fig:dist}
\end{figure}
Figure \ref{fig:dist} shows the scatter plot of the cohesive layers for $N_c=4$. For each cohesive layer, the x value is the volume fraction of its enriched node in the bottom layer of original material network, and the y value is its reciprocal length parameter $\tilde{v}$. The remaining number of active cohesive layers for $N=5, 7,$ and 9 are 2, 13, and 40. Since the network with $N=9$ has more enriched nodes, the mean volume fraction of the nodes is smaller, and more cohesive layers are deactivated. For all the cases, the mean reciprocal length parameter is around 0.05 $\text{mm}^{-1}$. This value is related to the characteristic length of the UD RVE, whose average diameter of the fibers is 25 mm.

\textit{Offline training time:} All training processes are parallelized using 10 processors. Here the computational times in stage II for networks with different depths are listed and compared. For $N=5$, networks with $N_c=2,4,6$ took 0.4, 0.5, 0.6 hours. For  $N=7$, the training times are  1.7, 2.0, 2.2 hours.  For $N=9$, the training times are 6.7, 7.9, 8.2 hours. Therefore, $N_c$ has less effect on the training time than $N$, indicating that most of the computational cost still locates at the original material network.

\subsection{Online extrapolation to interfacial failure analysis}\label{sec:online}
For the UD composite, three different loading directions are evaluated in the online stage: transverse tension and compression, transverse shear, and longitudinal shear. Here the case under longitudinal tension along the fiber direction is not presented, as there is almost no interfacial failure within the loading range. For each loading case, the overall strain component in the loading direction follows the path shown in Figure \ref{fig:ud} (b), while the overall stress components in the other directions are zero. For example, in transverse tension and compression, the loading conditions can be written as
\begin{equation}
\varepsilon_{11} = \begin{cases}
t & t \in [0,0.012)\\
0.024-t& t \in [0.012,0.030]
\end{cases},\quad \sigma_{22}=\sigma_{33}=\sigma_{23}=\sigma_{13}=\sigma_{12}=0.
\end{equation}
Accordingly, the absolute strain rate $|\dot{\varepsilon}_{11}|$ is 1.0 s$^{-1}$. The time step in both DNS and DMN is $\Delta t=1.0\times10^{-4}$. For simplicity, $\varepsilon$ and $\sigma$ are used to denote the overall strain and stress in this section. Snapshots of DNS are taken at $t=0.004,0.012,0.024,$ and 0.030 s to illustrate the evolution of RVE deformation during the loading.

{For all the online examples to appear in this section, the study shows that $N_c$ has little effect on the stress-strain predictions as long as $N_c\geq 2$, while the depth of the original material network $N$ is more prominent. Therefore,  $N_c$ is set to be 4 by default in the online DMNs. Ideally, a 3D material block can be fully covered by 3 cohesive layers if they are oriented in three orthogonal directions. In this regard, $N_c=4$ tends to be sufficient for the solution space to capture the interfacial effects, and more layers would not help to improve the accuracy too much. This statement on the choice of $N_c$ also applies to the particle-reinforced RVE shown in \ref{ap:ap3}. However, it will be investigated more coherently in the future.}

\begin{figure} [!t]
	\centering
	\graphicspath{{Figures/}}
	\subfigure[DNS snapshots of deformed RVE with plastic matrix. The displacements are magnified 10 times.]{\includegraphics[clip=true,trim = 3.5cm 7.5cm 6.5cm 6.8cm,width=0.98\textwidth]{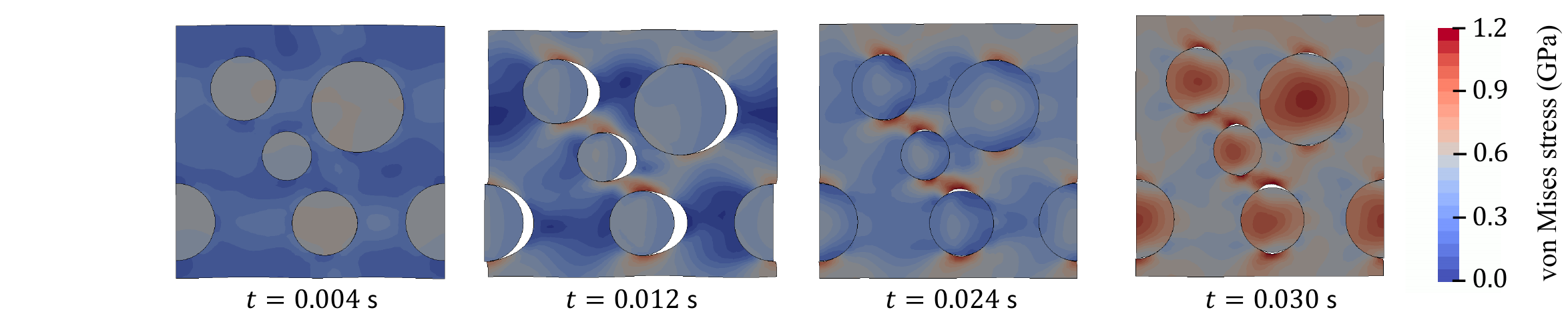}}
	\subfigure[$\sigma_{11}$ vs. $\varepsilon_{11}$, elastic matrix.]{\includegraphics[clip=true,trim = 0.0cm 0.0cm 1.0cm 0.5cm,width=0.44\textwidth]{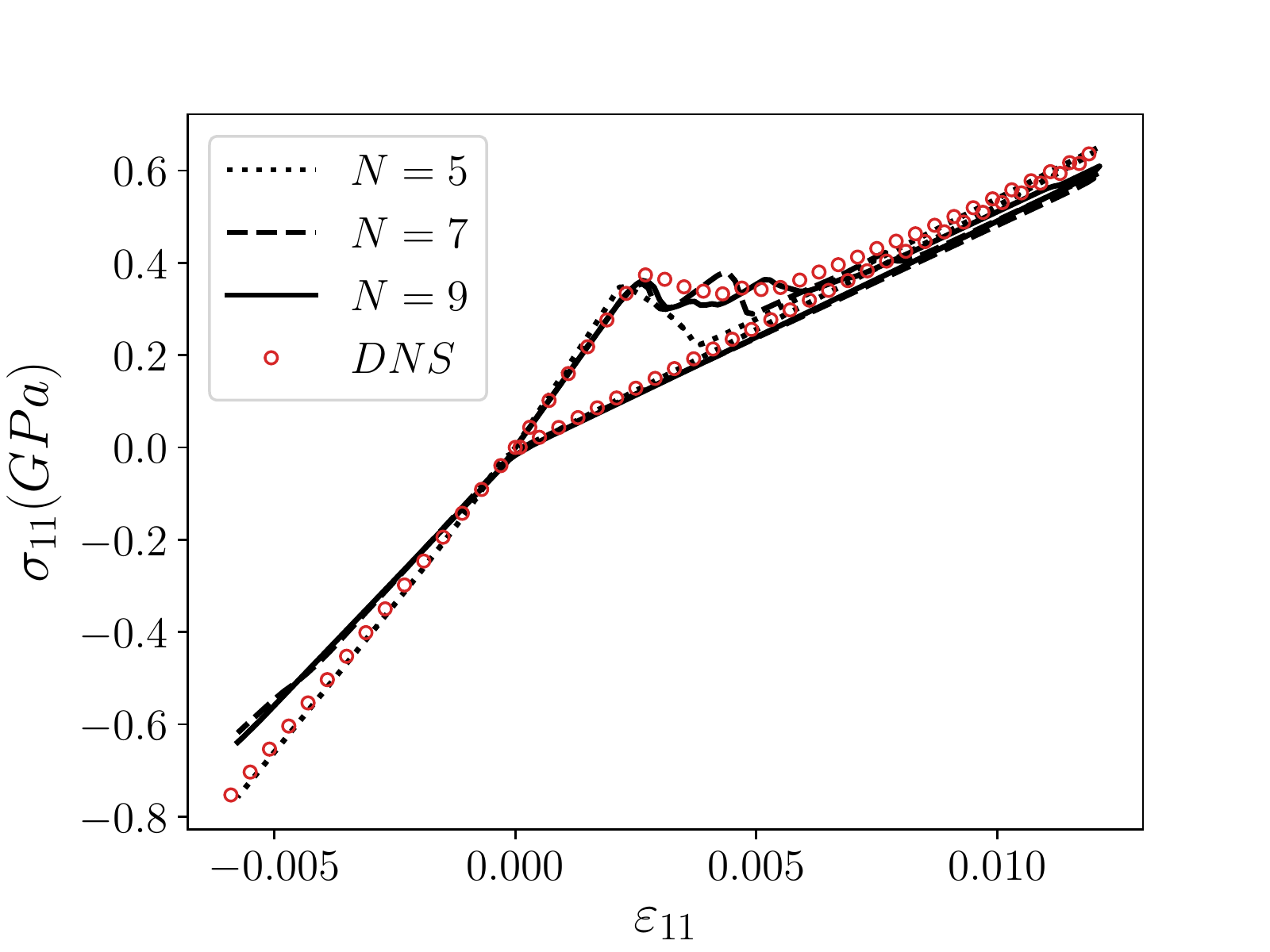}}
	\subfigure[$\sigma_{11}$ vs. $\varepsilon_{11}$, plastic matrix.]{\includegraphics[clip=true,trim = 0.0cm 0.0cm 1.0cm 0.5cm,width=0.44\textwidth]{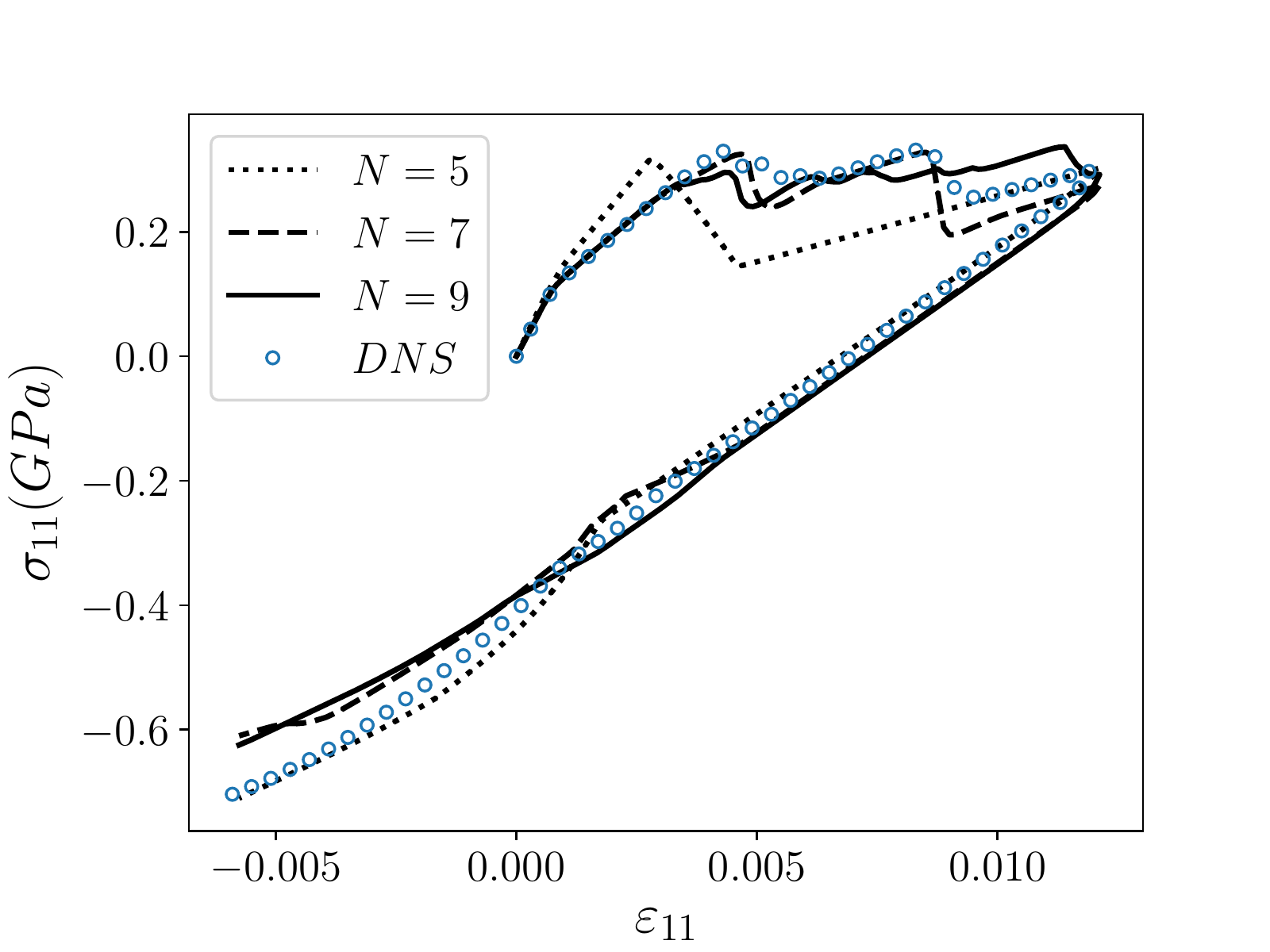}}
	\caption{Results of UD RVE with debonding interfaces under transverse tension and compression. Two cases are considered for the matrix phase: (b) Linear elasticity and (c) von Mises plasticity. Each cohesive network has 4 cohesive layers: $N_c=4$.}
	\label{fig:ud0}
\end{figure}

Figure \ref{fig:ud0} summarizes the results of transverse tension and compression. As shown in Figure \ref{fig:ud0} (a) for plastic matrix, no interfacial failure happens at $t=0.004$ s. Since the fiber phase is stiffer than the matrix phase, its average von Mises stress is higher. The interfaces normal to the loading direction start to debond at $t = 0.005$ s and reach the maximum opening at $t = 0.012$ s. Due to the debonding, less load transfers from the matrix to the fibers, and more stress concentration appears in the matrix phase. Around $t = 0.024$ s, closure of the interfaces happens. Interestingly, further compression results in debonding of the interfaces in the loading direction, which is related to the Poisson's effect. The corresponding stress-strain curves from DNS and material networks with $N=5,7,9$ and $N_c=4$ are provided in \ref{fig:ud0} (c). Stiffness degradation due to the failure of interfaces is observed during unloading. Other softening behaviors are also captured by DMN with cohesive layers. For $N=7$ and 9, the DMN predictions agree with the corresponding DNS results very well. 

The difference between transverse tension and compression is more evident on the plot for elastic matrix shown in Figure \ref{fig:ud0} (b). Since there is no plasticity in the matrix, during unloading, the stress-strain curve almost goes back to the origin with a degraded stiffness. However, in compression, the interfaces are in contact, so that the stiffness returns to a high value and the response is linear elastic without any failure. It appears that the DMN slightly underestimates the stiffness under compression. Nevertheless, these physical phenomena related to the interface debonding and closure are all captured by DMNs.

\begin{figure} [!t]
	\centering
	\graphicspath{{Figures/}}
	\subfigure[DNS snapshots of deformed RVE with plastic matrix. The displacements are magnified 10 times.]{\includegraphics[clip=true,trim = 3.5cm 7.2cm 6.5cm 6.8cm,width=0.98\textwidth]{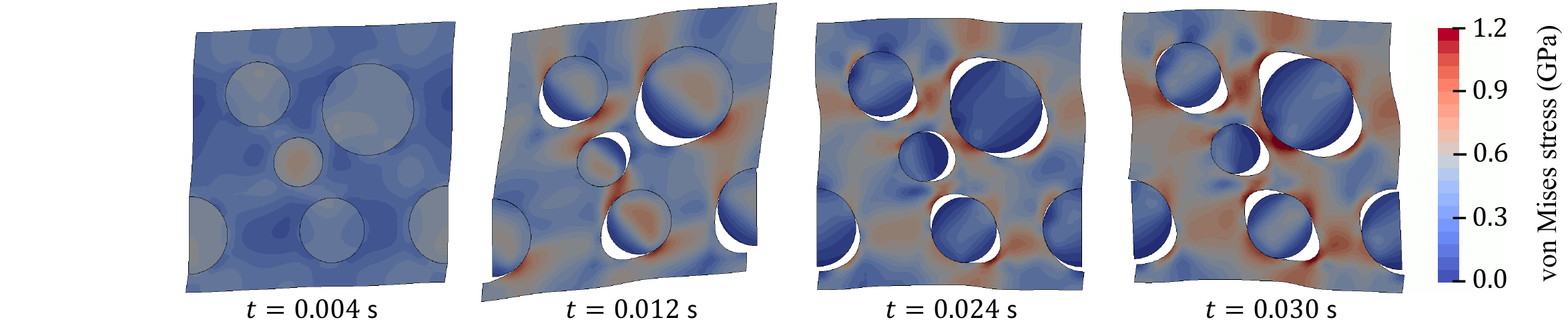}}
	\subfigure[$\sigma_{12}$ vs. $\varepsilon_{12}$, elastic matrix.]{\includegraphics[clip=true,trim = 0.0cm 0.0cm 1.0cm 0.5cm,width=0.44\textwidth]{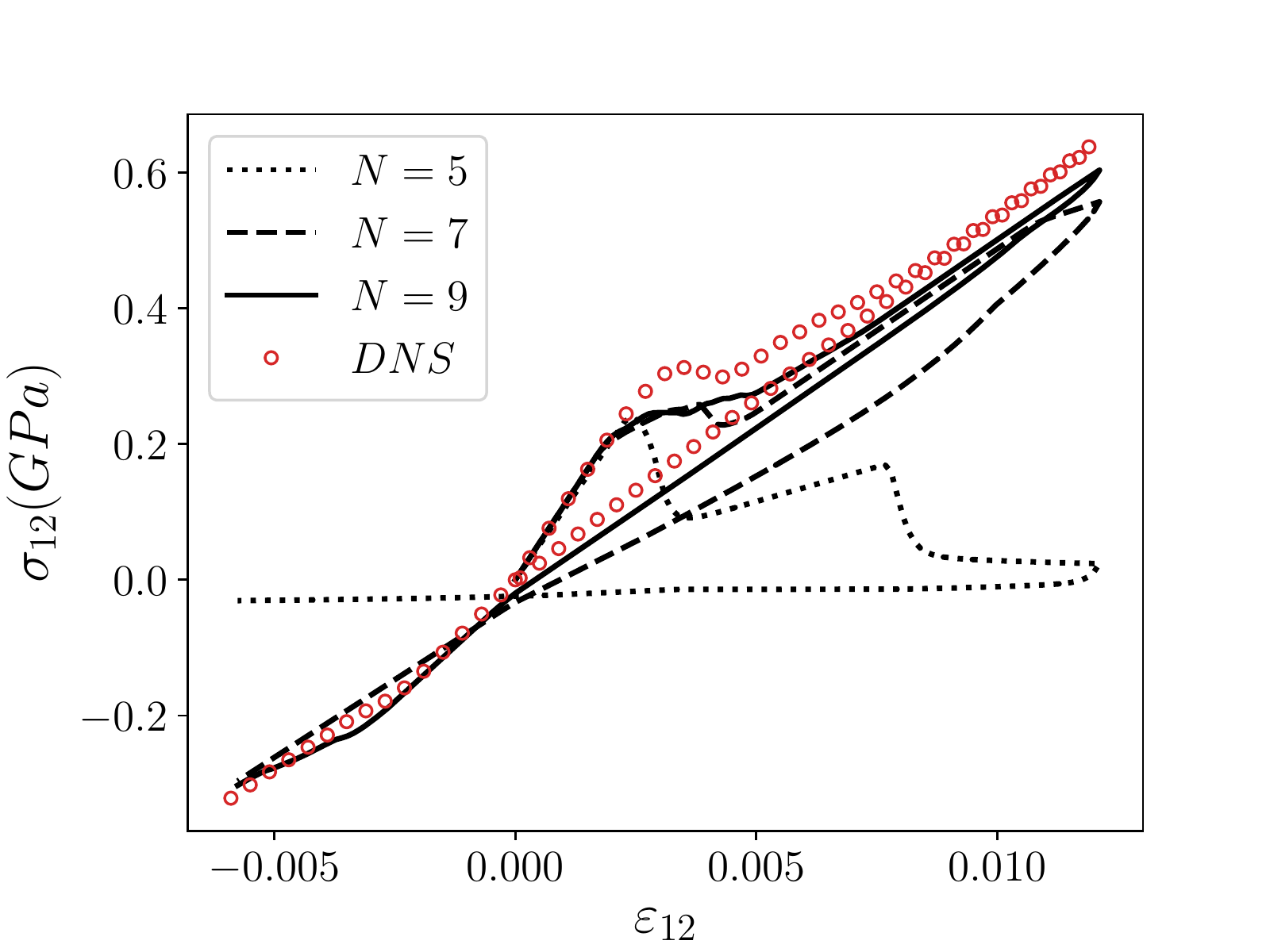}}
	\subfigure[$\sigma_{12}$ vs. $\varepsilon_{12}$, plastic matrix.]{\includegraphics[clip=true,trim = 0.0cm 0.0cm 1.0cm 0.5cm,width=0.44\textwidth]{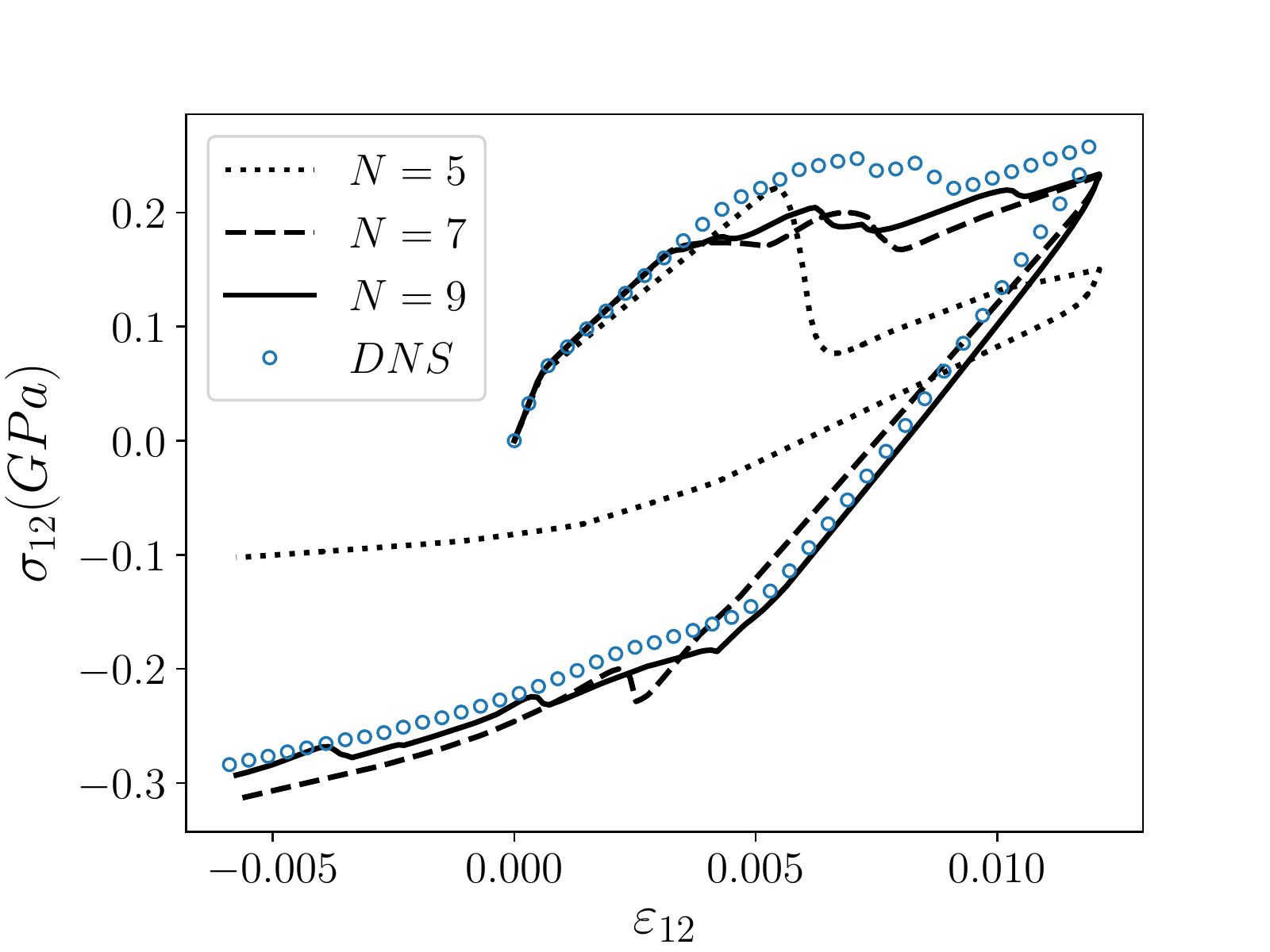}}
	\caption{Results of UD RVE with debonding interfaces under transverse shear. Two cases are considered for the matrix phase: (b) Linear elasticity and (c) von Mises plasticity. Each cohesive network has 4 cohesive layers: $N_c=4$.}
	\label{fig:ud1}
\end{figure}

Figure \ref{fig:ud1} summarizes the results from transverse shear loading.  The DNS snapshot at $t=0.012$ s shows that the debonding of interfaces is driven by the normal traction. After reversing the loading direction, previously debonded interfaces are closed, while new surfaces are created at different locations. As one can see from Figure \ref{fig:ud1} (b), the reverse-loading part of stress-strain curve still undergoes some softening effect during the plastic yielding, which is consistent with those findings in the DNS snapshots. For either case with elastic or plastic matrix, the DMN with $N=9$ is able to capture the overall stress-strain response, while the amount of interfacial failure is over-predicted. The network with $N=5$ has a bad performance in predicting the transverse shear behavior. Hence, representing the fiber phase with only one node in DMN (see Figure \ref{fig:udTree} (a)) is not sufficient for the debonding analysis of UD RVE.

\begin{figure} [!t]
	\centering
	\graphicspath{{Figures/}}
	\subfigure[DNS snapshots of deformed RVE with plastic matrix. The displacements are magnified 10 times.]{\includegraphics[clip=true,trim = 3.5cm 7.2cm 6.5cm 6.8cm,width=0.98\textwidth]{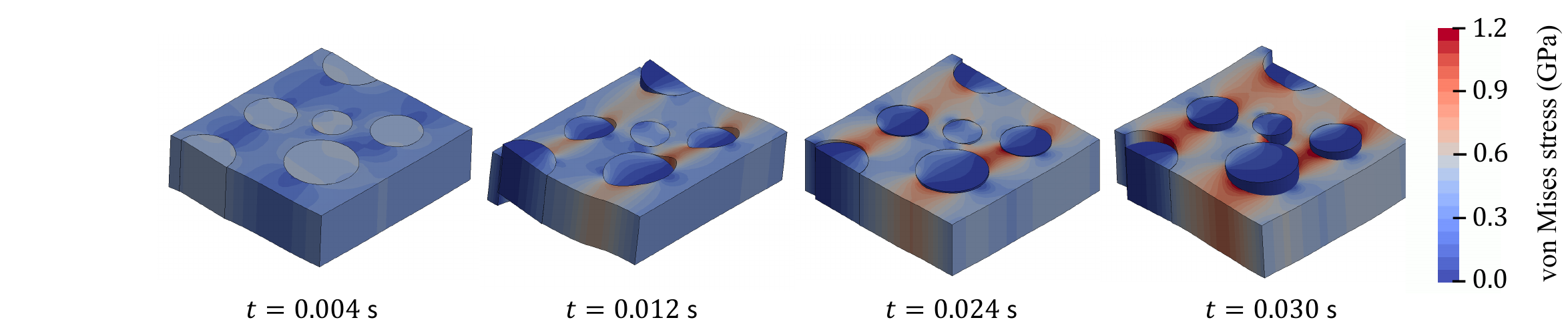}}
	\subfigure[$\sigma_{23}$ vs. $\varepsilon_{23}$, elastic matrix.]{\includegraphics[clip=true,trim = 0.0cm 0.0cm 1.0cm 0.5cm,width=0.44\textwidth]{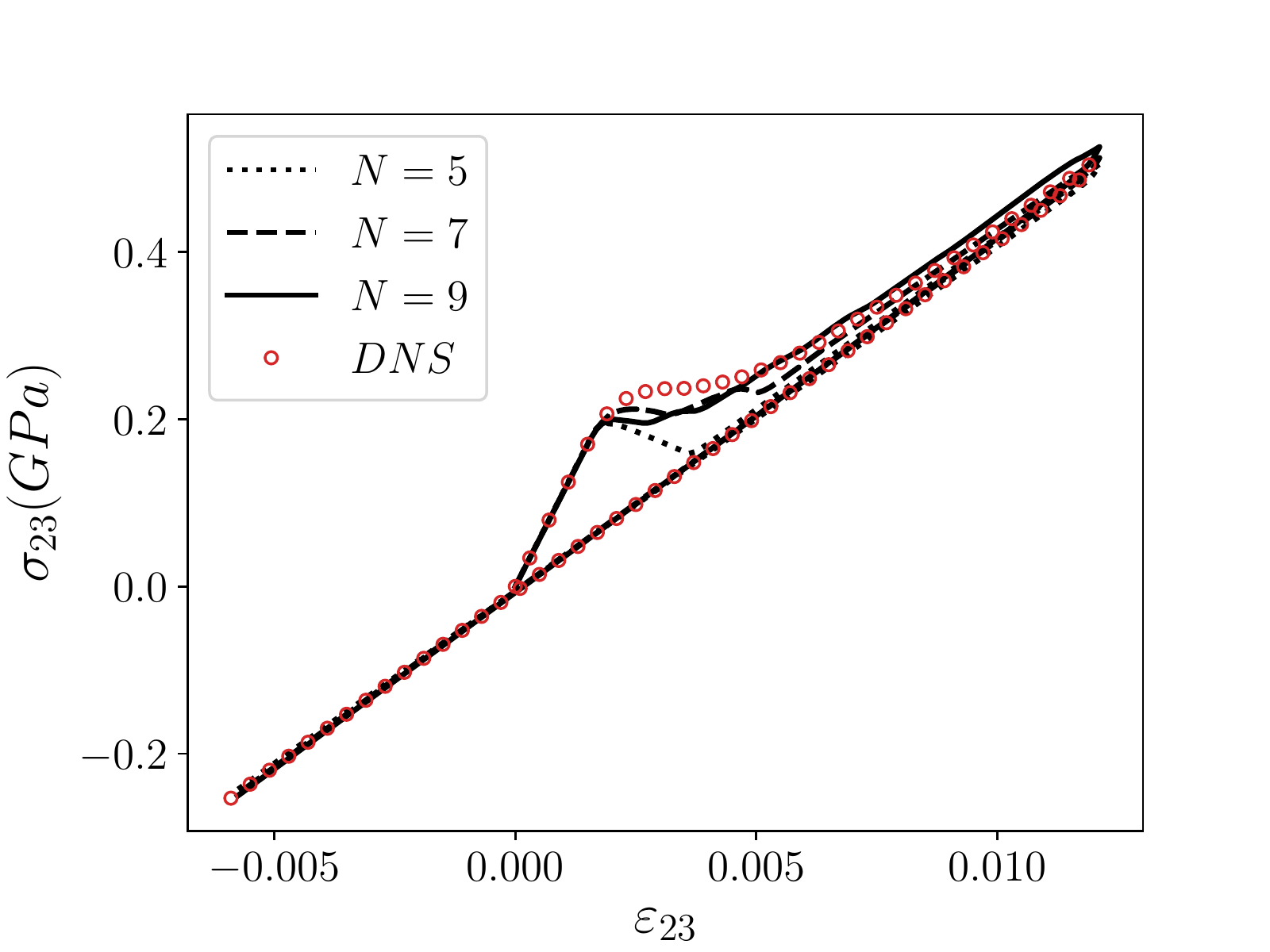}}
	\subfigure[$\sigma_{23}$ vs. $\varepsilon_{23}$, plastic matrix.]{\includegraphics[clip=true,trim = 0.0cm 0.0cm 1.0cm 0.5cm,width=0.44\textwidth]{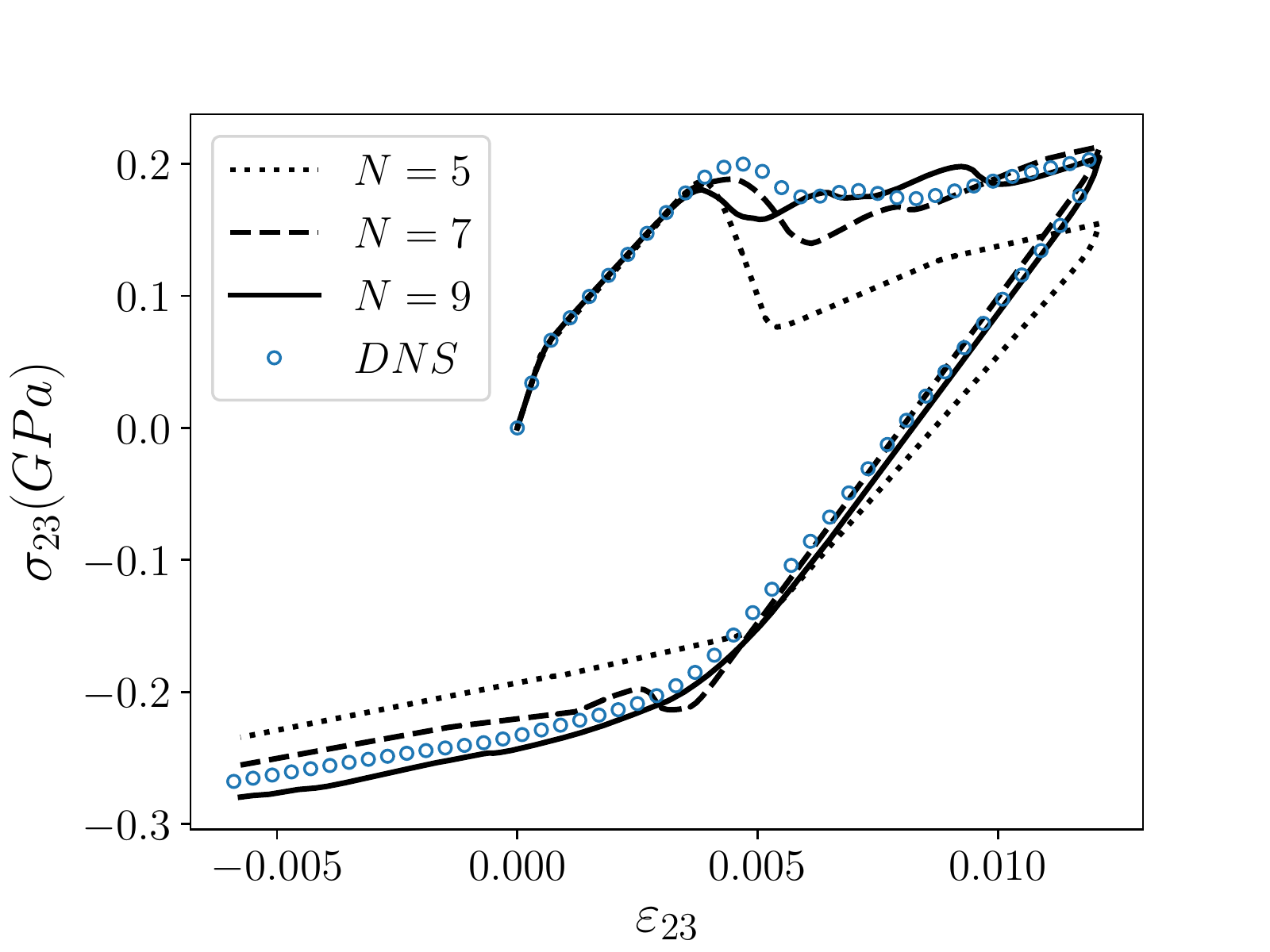}}
	\caption{Results of UD RVE with debonding interfaces under longitudinal shear. Two cases are considered for the matrix phase: (b) Linear elasticity and (c) von Mises plasticity. Each cohesive network has 4 cohesive layers: $N_c=4$.}
	\label{fig:ud2}
\end{figure}
Figure \ref{fig:ud2} summarizes the results from longitudinal shear loading. The snapshots show that the interfacial failure is driven by the sliding deformation, which is different from the previous two loading cases. Moreover, after reversing the loading direction at $t=0.012$ s, the sliding locations still stay the same, indicating that the amount of newly failed interfaces is small. This is reflected in the corresponding stress-strain curve plotted in Figure \ref{fig:ud2} (c), where the overall RVE behavior is elasto-plastic with negligible softening effect. Similarly, in Figure \ref{fig:ud2} (b) for elastic matrix, the stress-strain curve after reversing loading direction is almost linear elastic, and passes the origin. For both cases, the DMN predictions with $N=7$ and 9 agree with the DNS results very well.

\textit{Online prediction time:} The DMN with cohesive layers learns an efficient reduced-order model of the DNS RVE with deformable interfaces. Both DNS and DMN models have 300 loading steps. A typical DNS with plastic matrix took 5.6 h on 10 processors. Yet it is noteworthy that the original DNS with perfectly bonded interfaces (or no cohesive element) only took 1.6 h on 10 CPUs, implying that the softening interfaces consume more iterations in the implicit analysis. In comparison, DMNs with cohesive layers took 1.4, 5.8 and 33.3 s on one CPU for with $N=5, 7$ and 9, respectively. In terms of the CPU time, the enriched material network with $N=9$ and $N_c=4$ is more than 6000 times faster than the DNS.

\subsection{Predictions on local properties}
Local properties can also be predicted by the DMN. In the online stage, each active node in the bottom layer of DMN is treated as an independent microscale material,  representing either the fiber or the matrix phase in the UD composite. Meanwhile, each cohesive layer can be regarded as a part of the fiber-matrix interfaces, and its effective area is defined in Eq. (\ref{eq:effectiveA}), which is related to the reciprocal length parameter. With the volume fractions of the bottom-layer nodes and the effective areas of the cohesive layers, one can reproduce the probability distribution of local fields in the 3D bulk material and the 2D interfaces, respectively. In this section, the DMN with $N=9, N_c=4$ is taken as an example, which has 60 active bottom-layer nodes and 40 active cohesive layers after the training. While for the DNS, there are 63720 solid elements and 4320 cohesive elements.
\begin{figure} [!t]
	\centering
	\graphicspath{{Figures/}}
	\subfigure[von Mises stress in fiber and matrix phases.]{\includegraphics[clip=true,trim = 0.0cm 0.0cm 1.0cm 0.5cm,width=0.44\textwidth]{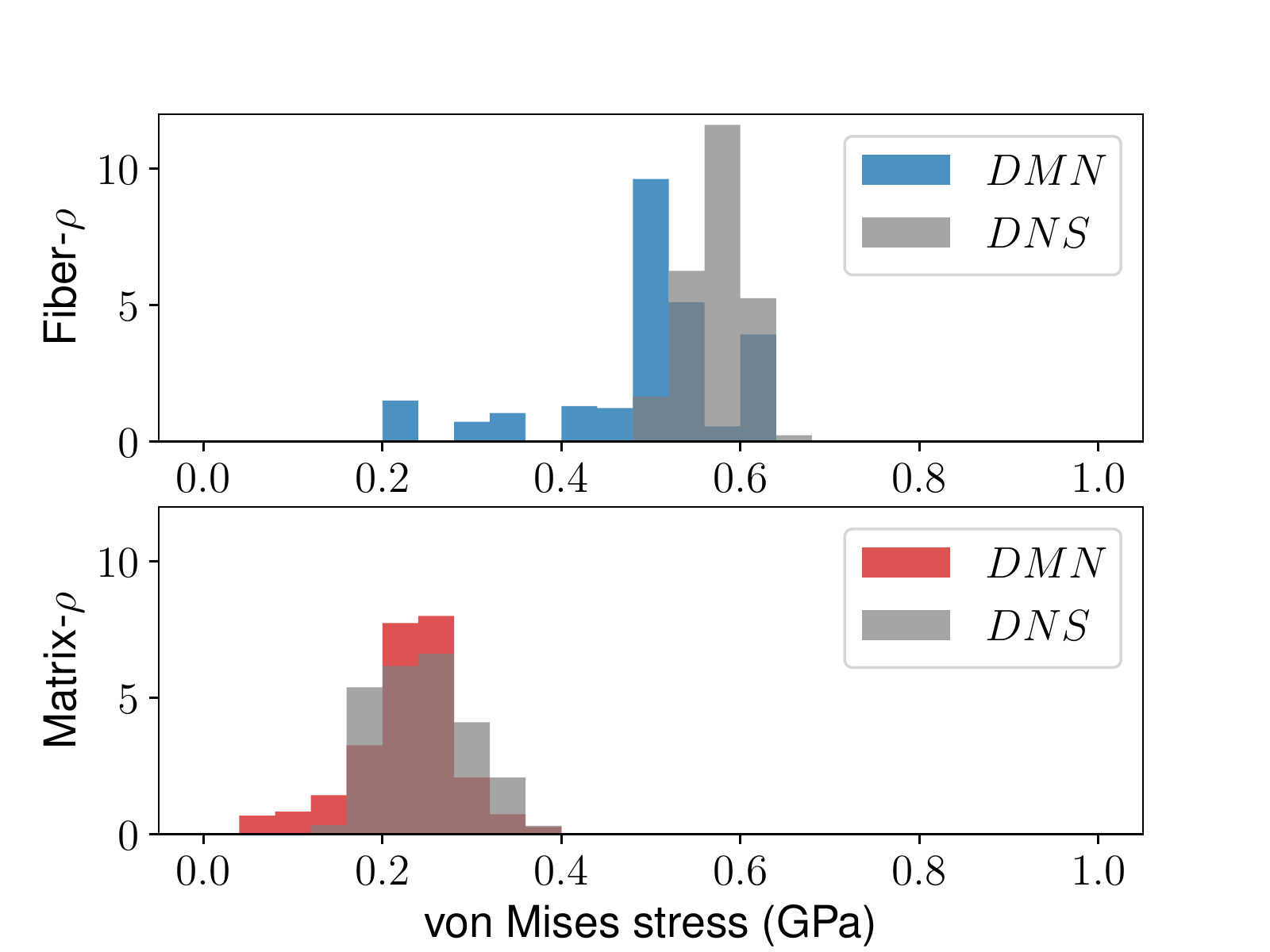}}
    \subfigure[Normal traction at interface.]{\includegraphics[clip=true,trim = 8.8cm 3.4cm 9.7cm 3.4cm,width=0.44\textwidth]{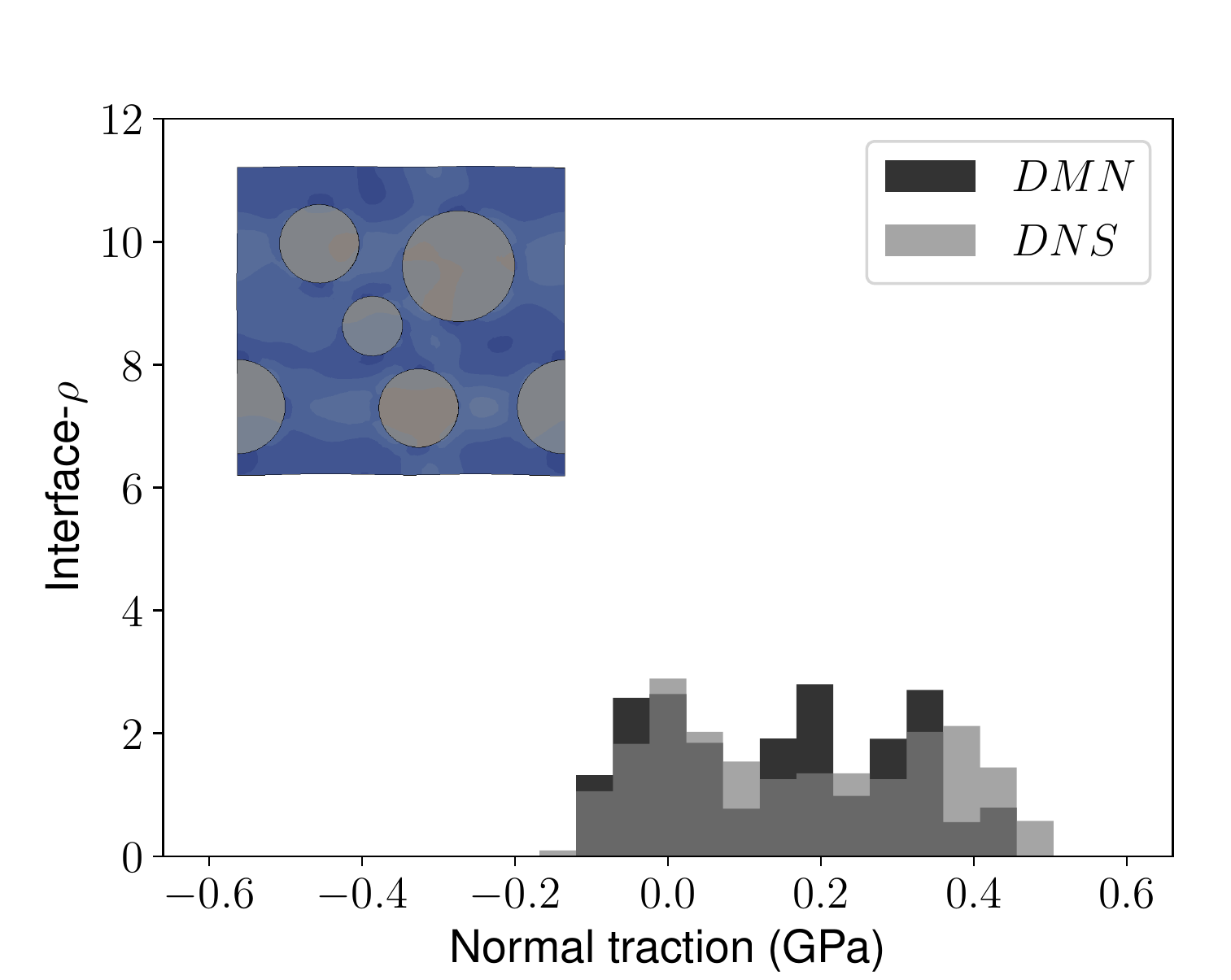}}
	\caption{Local fields predicted by DMN and DNS under uniaxial tension at $t = 0.004$ s and $\varepsilon_{11}=0.004$.}
	\label{fig:ud_local0}
\end{figure}
\begin{figure} [!t]
	\centering
	\graphicspath{{Figures/}}
	\subfigure[von Mises stress in fiber and matrix phases.]{\includegraphics[clip=true,trim = 0.0cm 0.0cm 1.0cm 0.5cm,width=0.44\textwidth]{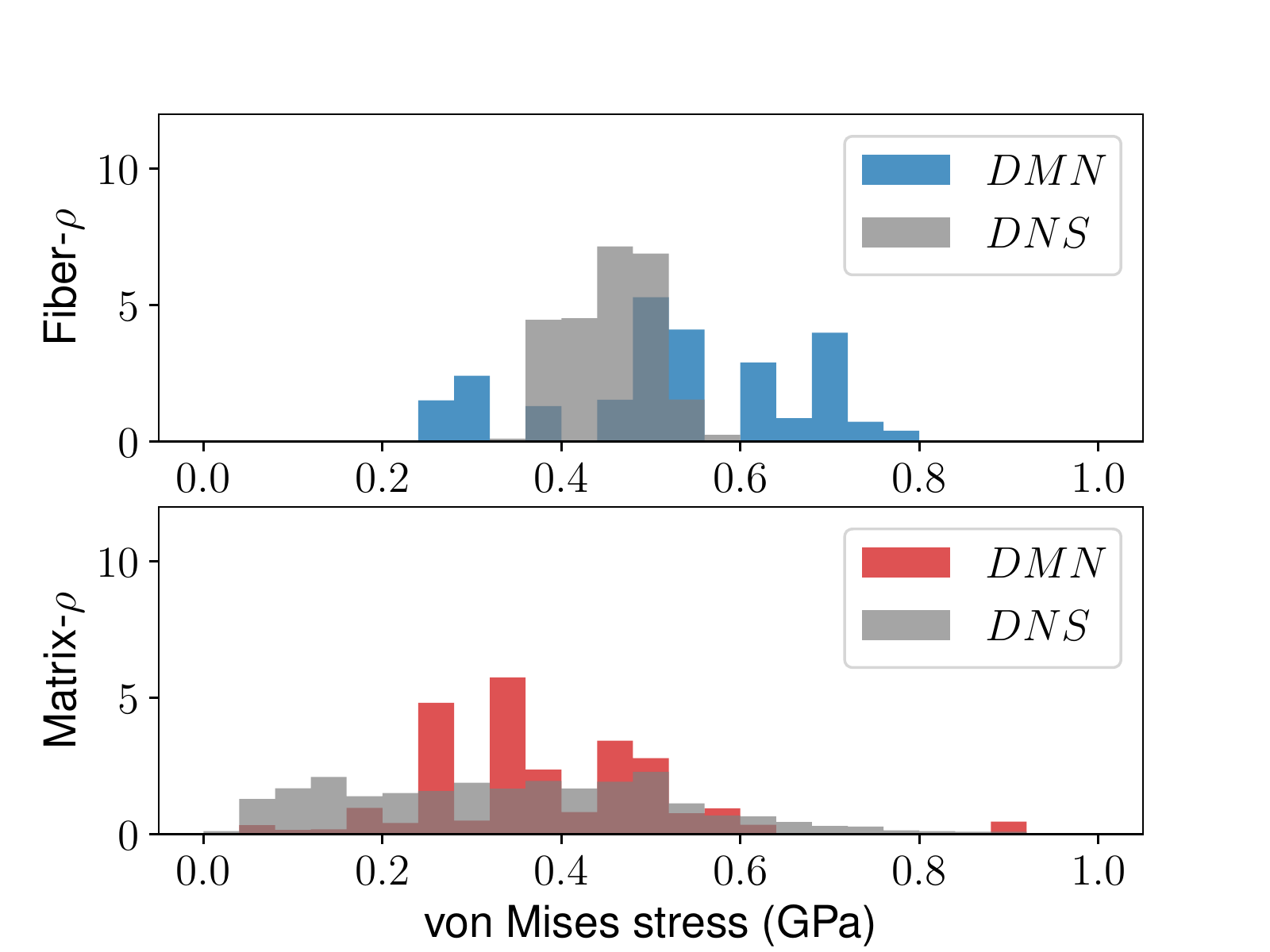}}
	\subfigure[Normal traction at interface.]{\includegraphics[clip=true,trim = 8.8cm 3.4cm 9.7cm 3.4cm,width=0.44\textwidth]{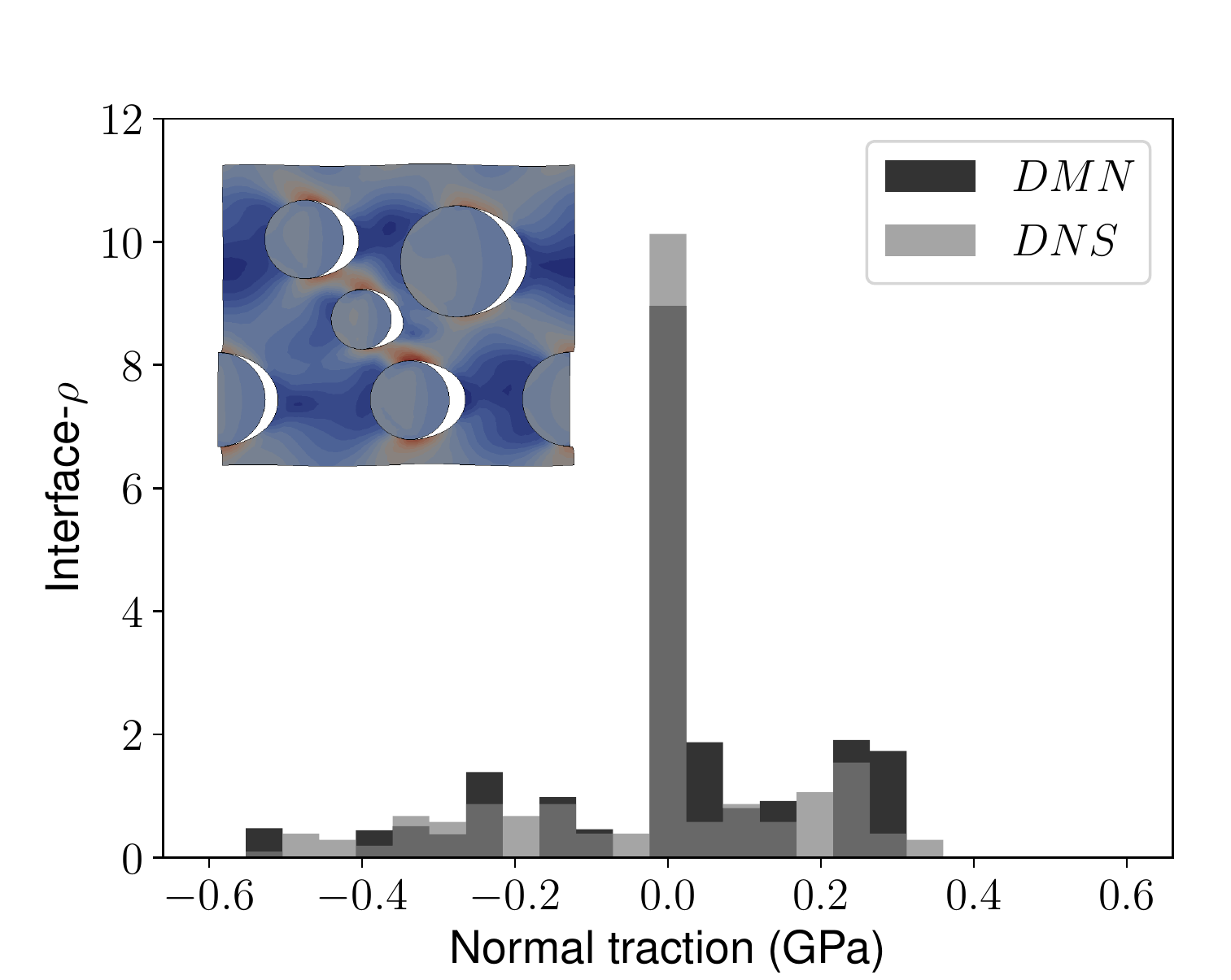}}
	\caption{Local fields predicted by DMN and DNS under uniaxial tension at $t = 0.012$ s and $\varepsilon_{11}=0.012$.}
	\label{fig:ud_local1}
\end{figure}

Figure \ref{fig:ud_local0} shows the distributions of various local fields predicted by DMN and DNS with plastic matrix under uniaxial tension at $t=0.004$ s, while the strain in the loading direction is $\varepsilon_{11}=0.004$. The corresponding global stress-strain curves are given in Figure \ref{fig:ud0} (c). For the matrix and fiber phases, the distributions of von Mises stress are compared in Figure \ref{fig:ud_local0} (a). Since most of the interfaces are still bonded, the average von Mises stress of the fiber phase is more than two times larger than that of the matrix phase. For the interfaces, the normal traction ranges from -0.15 GPa to 0.45 GPa, which is less than the critical effective traction $\sigma_c=0.5$ GPa. Furthermore, the distributions of local fields from DMN and DNS at $t=0.012$ s and $\varepsilon_{11}=0.012$ are presented in Figure \ref{fig:ud_local1}.  Due to debonding, the profile of the normal traction in Figure \ref{fig:ud_local1} (b) has a large density at $0$ GPa, which corresponds to those failed interfaces in DNS or cohesive layers in DMN. As expected, the maximum normal traction is still less than the critical effective traction 0.5 GPa, while the compressive traction reaches -0.55 GPa. The distribution of von Mises stress in the matrix phase becomes more dispersed, and the average stress is close to the one in the fiber phase. Although the DMN has much less DOF, it gives a good estimation on the distributions of local fields comparing to the DNS reference.

\section{Conclusions and future work} \label{sec:conclusion}
A new architecture of DMN with cohesive layers is proposed to consider deformable interfaces inside a heterogeneous material. In the physics-based cohesive building block, the activation of a cohesive layer is governed by a reciprocal length parameter, which also defines the effective thickness of the building block in its normal direction. Due to the existence of analytical solutions, the new material network can still be trained using gradient-based optimization methods and the backpropagation algorithm. Moreover, a two-stage training strategy is utilized to learn the fitting parameters in the original material network and the cohesive networks separately. 

The predictive capability of the method is demonstrated through the debonding analysis of the UD composite under various loading conditions, including transverse tension and compression, transverse shear, and longitudinal shear. The behavior of the fiber-matrix interface is described by a mixed-mode irreversible cohesive law. Though the material network is trained with linear elastic data, it can accurately capture the nonlinear RVE responses both locally and globally with complexity coming from the plastic matrix and the softening interfaces.  Involving less DOF, the enriched material network with $N=9$ and $N_c=4$ is more than 6000 times faster than the DNS in terms of the CPU time.

The proposed method opens new possibilities in materials engineering and design. First, DMN offers an efficient way of realizing concurrent multiscale simulations. With the addition of cohesive layers, it provides an effective data-driven material model for the interfacial failure analysis of a large-scale heterogeneous structure. Since measuring interface properties directly from the experiments can sometimes be challenging, DMN can also be used for inverse prediction of those unknown properties. Finally, with the transfer learning technique proposed in \cite{liu2019transfer}, the unified microstructural databases can potentially be created for the design of various modern material systems with nontrivial interfacial effects.

\section*{Acknowledgment}
Z. Liu would like to acknowledge Dr. C.T. Wu, Dr. Yong Guo, Dr. Tadashi Naito, and Tianyu Huang for helpful discussions. Z. Liu warmly thanks Dr. John O. Hallquist for his support to this research. This is a preprint version of an article published in \text{Computer Methods in Applied Mechanics and Engineering}. The final authenticated version is available online at: \url{https://doi.org/10.1016/j.cma.2020.112913}.

\appendix
\section{Application to particle-reinforced composite}\label{ap:ap3}
\begin{table}[!t]
	\captionabove{Training results of the particle-reinforced RVE at stages I and II. Average training error $\bar{e}^{tr}$, average test error $\bar{e}^{tr}$ and maximum test error are provided for each DMN. In stage II, the number of cohesive layers in each cohesive network is $N_c=4$.} 
	\centering 
	\label{table:np} 
	{\tabulinesep=1.0mm
		\begin{tabu}{c | c c c |c c c} 
			\hline 
			&&\textbf{Stage I}&&&\textbf{Stage II}&\\
			&Training $\bar{e}^{tr}$ & Test $\bar{e}^{te}$ & Maximum $e_s^{te}$ &Training $\bar{e}^{tr}$ & Test $\bar{e}^{te}$ & Maximum $e_s^{te}$\\
			\hline
			$N=4$&7.61\%&7.79\%&17.9\%&8.12\%&8.40\%&17.9\%\\
			$N=6$&1.34\%&1.39\%&4.46\%&1.47\%&1.49\%&2.81\%\\
			$N=8$&0.53\%&0.59\%&2.41\%&0.64\%&0.67\%&2.47\% \\
			\hline
	\end{tabu}}
\end{table}
\begin{figure}[!t]
	\centering
	\graphicspath{{Figures/}}
	\subfigure[DNS snapshots at $\varepsilon_{11}=0.012$, magnified 20 times. ]{\includegraphics[clip=true,trim = 12.0cm 5.5cm 11.0cm 3.0cm,width=0.44\textwidth]{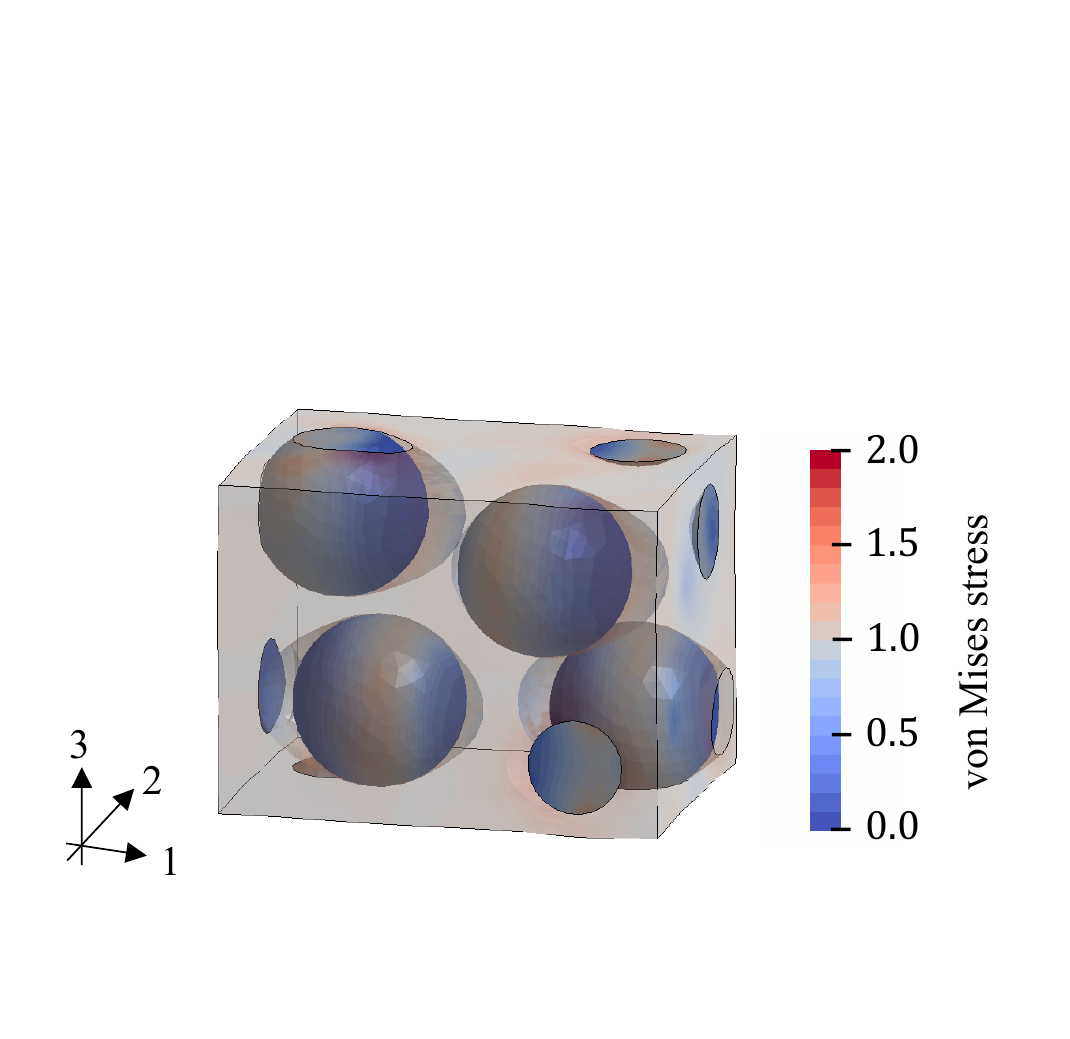}}
	\subfigure[$\sigma_{11}$ vs. $\varepsilon_{11}$, plastic matrix.]{\includegraphics[clip=true,trim = 0.0cm 0.0cm 1.0cm 0.5cm,width=0.44\textwidth]{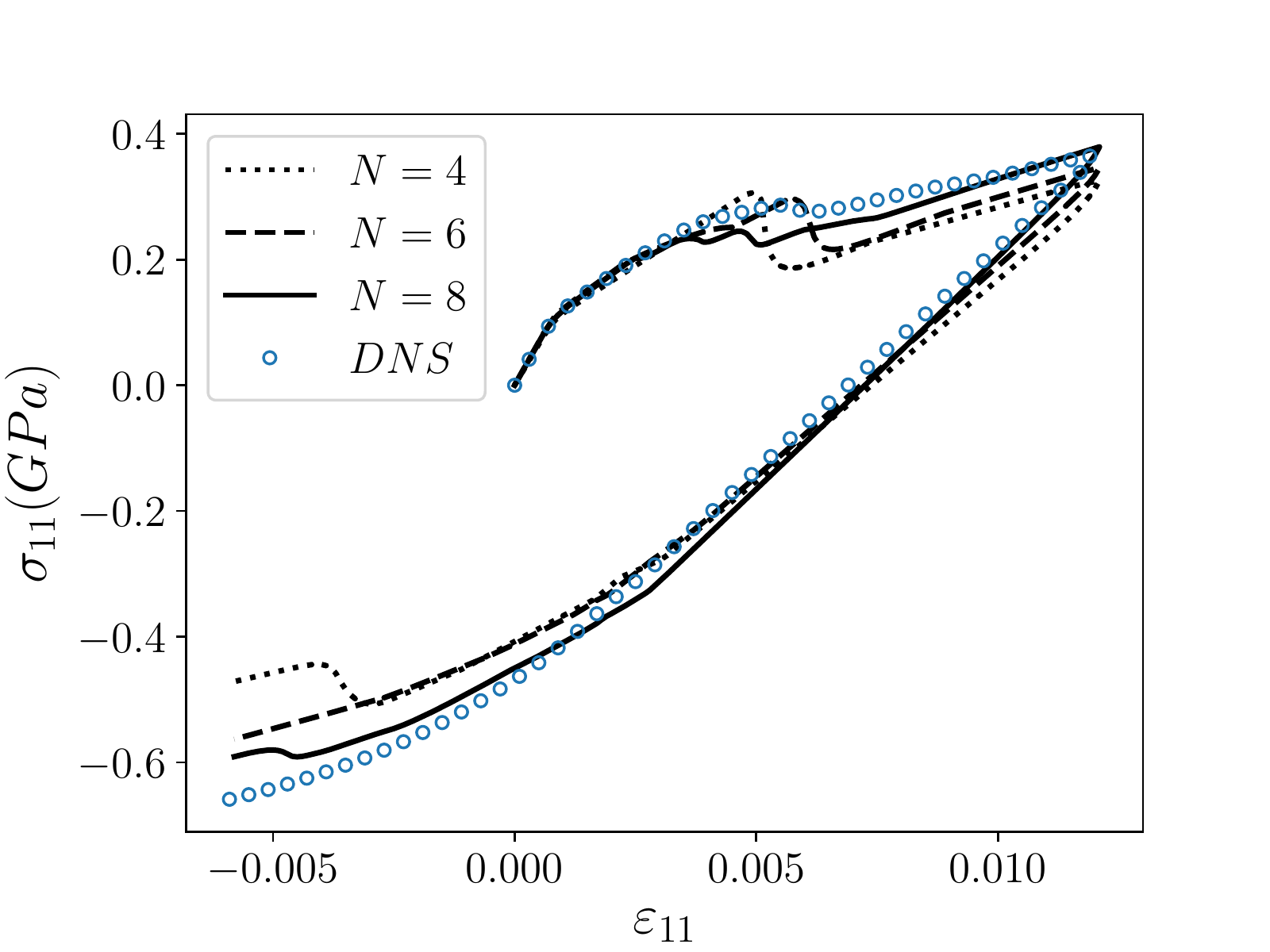}}
	\caption{Particle-reinforced RVE: (a) The volume fraction of the particle phase is 22.6\% while the FE model has 84693 nodes, 59628 10-node tetrahedron elements, and 9647 6-node zero-thickness cohesive elements; (b) Stress-strain curves under uniaxial tension and compression.}
	\label{fig:np}
\end{figure}
Other than the UD composite presented in the main content, the deep material network with cohesive layers is applied to a particle-reinforced composite. Volume fraction of the particle phase is 22.6\%. In terms of the DNS, the FE model has 84693 nodes, 59628 10-node tetrahedron elements, and 9647 6-node zero-thickness cohesive elements. The same material properties in Table \ref{table:materialpara} are assigned to the new RVE, while its particle phase shares the same properties with the fiber phase of the UD composite. The characteristic length $L$ is set to be 5.0 mm, which is same as the diameter of the particles.

In training stage I, three material networks are evaluated with different depths $N=4$, 6, and 8, which are trained for 20000, 20000, and 40000 epochs, respectively. In training stage II, the number of cohesive layers in each cohesive network is $N_c=4$, and all three enriched DMNs are trained for 5000 epochs. The training and test errors of both stages are listed in Table \ref{table:np}. With depth $N=8$, the average training and test errors of both stages I and II are reduced to be less than 1\%.

The results under uniaxial tension and compression are provided in Figure \ref{fig:np}. The loading history of $\varepsilon_{11}$ is shown in Figure \ref{fig:ud} (b), while the RVE is stress-free in the other directions. As shown in the DNS snapshot at $\varepsilon_{11}=0.012$, the particle-matrix interfaces are debonded in the loading direction. The matrix phase is made partially transparent to show the inner separated surfaces. Furthermore, the stress-strain curves predicted by DNS and DMNs are plotted in Figure \ref{fig:np} (b). Due to the contact of debonded interfaces, the particle-reinforced RVE is stiffer under compression. A good agreement is observed between DNS and DMNs with $N=6$ and 8.

\section{Elementary rotation matrices}\label{ap:ap0}
In small-strain formulation, the elementary rotation matrices shown in Eq. (\ref{eq:smallR}) are given by
\begin{equation}
\text{X}_{(1,1)}=1,\quad\textbf{X}_{([2,3,4],[2,3,4])}(\alpha)=\textbf{r}^p(\alpha),\quad \textbf{X}_{([5,6],[5,6])}(\alpha)=\textbf{r}^v(\alpha);
\end{equation}
\begin{equation*}
\text{Y}_{(2,2)}=1,\quad\textbf{Y}_{([1,3,5],[1,3,5])}(\beta)=\textbf{r}^p(-\beta),\quad \textbf{Y}_{([4,6],[4,6])}(\beta)=\textbf{r}^v(-\beta);
\end{equation*}
\begin{equation*}
\text{Z}_{(3,3)}=1,\quad\textbf{Z}_{([1,2,6],[1,2,6])}(\gamma)=\textbf{r}^p(\gamma),\quad \textbf{Z}_{([4,5],[4,5])}(\gamma)=\textbf{r}^v(\gamma).
\end{equation*}
The in-plane and output-plane rotation matrices $\textbf{r}^p$ and $\textbf{r}^v$ for an arbitrary angle $\theta$ are defined in Mandel notation as
\begin{equation}
\textbf{r}^p(\theta)=\begin{Bmatrix}
\cos^2\theta&\sin^2 \theta&\sqrt{2}\sin\theta\cos\theta\\
\sin^2 \theta&\cos^2\theta&-\sqrt{2}\sin\theta\cos\theta\\
-\sqrt{2}\sin\theta\cos\theta&\sqrt{2}\sin\theta\cos\theta&\cos^2\theta-\sin^2\theta\\
\end{Bmatrix},
\quad\textbf{r}^v(\theta)=\begin{Bmatrix}
\cos \theta&-\sin\theta\\
\sin\theta&\cos \theta\\
\end{Bmatrix}.
\end{equation}

\section{Finite-strain solution of cohesive building block} \label{ap:ap1}
For finite-strain problem, the deformation gradient $\utilde{\textbf{F}}$ and first Piola-Kirchhoff stress $\utilde{\textbf{P}}$ are chosen as the strain and stress measures, respectively. For 3-dimensional problems, $\utilde{\textbf{F}}$ and $\utilde{\textbf{P}}$ are vectorized to simplify the computation:
\begin{equation}\label{eq:finiteFP}
\textbf{F} = \{\utilde{F}_{11},\utilde{F}_{22},\utilde{F}_{33},\utilde{F}_{23},\utilde{F}_{32},\utilde{F}_{13},\utilde{F}_{31},\utilde{F}_{12},\utilde{F}_{21}\}^T=\{F_{1},F_{2},F_{3},...,F_{9}\}^T,
\end{equation}
\begin{equation*}
\textbf{P} = \{\utilde{P}_{11},\utilde{P}_{22},\utilde{P}_{33},\utilde{P}_{23},\utilde{P}_{32},\utilde{P}_{13},\utilde{P}_{31},\utilde{P}_{12},\utilde{P}_{21}\}^T=\{P_{1},P_{2},P_{3},...,P_{9}\}^T.
\end{equation*}
In the generic cohesive building block, the constitutive relation of material 0 in finite-strain formulation can be written as
\begin{equation}
\Delta \textbf{F}^0=\textbf{B}^0\Delta \textbf{P}^0 +\delta \textbf{F}^0,
\end{equation}
where $\textbf{B}^0$ and $\delta \textbf{F}^0$ are the finite-strain compliance matrices and residual deformation gradient, respectively. The overall constitutive relation of the building block is
\begin{equation}
\Delta \textbf{F}=\textbf{B}\Delta \textbf{P} +\delta \textbf{F}.
\end{equation}
Based on the equilibrium condition, $\textbf{B}$ and $\delta \textbf{F}$ are derived to be
\begin{equation}
{\textbf{B}} ={\textbf{B}}^0+\tilde{v}\textbf{R}^f(\tilde{\alpha}, \tilde{\beta},\tilde{\gamma})\tilde{\textbf{G}}^f\left[\textbf{R}^{f}(\tilde{\alpha}, \tilde{\beta}, \tilde{\gamma})\right]^{-1}
\end{equation}
and
\begin{equation}
\delta{\textbf{F}} =\delta\textbf{F}^0+\tilde{v}\textbf{R}^f(\tilde{\alpha}, \tilde{\beta},\tilde{\gamma})\delta\tilde{\textbf{d}}^f.
\end{equation}
The nonzero terms in $\tilde{\textbf{G}}^f$ and $\delta\tilde{\textbf{d}}^f$ are
\begin{equation}
\tilde{{G}}^f_{([3,4,6],[3,4,6])}= \begin{Bmatrix}
{G}_{nn}&{G}_{ns} &{G}_{nt} \\
&{G}_{ss}&{G}_{st}\\
sym & & {G}_{tt}
\end{Bmatrix},\quad
\delta \tilde{{d}}^f_{([3,4,6])} = \begin{Bmatrix}
\delta d_n\\
\delta d_s\\
\delta d_t\\
\end{Bmatrix}.
\end{equation}
The compliance matrix ${\textbf{G}}$ and residual displacement $\delta\textbf{d}$ of the cohesive layer are defined in Eq. (\ref{eq:cohe}). The definition of the finite-strain rotation matrix $\textbf{R}^f$ can be found in \cite{liu2019exploring}, which can also be decomposed into three elementary rotation matrices.

\section{Tangent stiffness matrix and viscous regularization} \label{ap:ap2}
Since both DNS and DMN are based on implicit analysis, the tangent stiffness matrix of the cohesive law needs to be derived, which is defined as
\begin{equation}
\textbf{K} = \dfrac{\partial \textbf{t}}{\partial \textbf{d}}.
\end{equation}
Again, the tensile and compressive cases are treated differently:
\begin{enumerate}
	\item For tensile case $d_n\geq0$, the components of $\textbf{t}$ are
	\begin{equation}
	t_n = \dfrac{t_m}{d_m}{d_n}, \quad t_s =\dfrac{\beta^2t_m}{d_m}{d_s},\quad t_t =\dfrac{\beta^2t_m}{d_m}{d_t}.
	\end{equation}
	The independent components in $\textbf{K}$ are
	\begin{equation}
	K_{nn} = \dfrac{t_m}{d_m}+{{d_n}^2}\Delta,\quad
	K_{ns} = \beta^2{d_n}{d_s}\Delta,\quad K_{nt} = \beta^2{d_n}{d_t}\Delta,
	\end{equation}
	\begin{equation*}
	K_{ss} = \dfrac{\beta^2t_m}{d_m}+\beta^4{d_s}^2\Delta,\quad K_{ss} = \dfrac{\beta^2t_m}{d_m}+\beta^4{d_t}^2\Delta,\quad K_{st} = \beta^4{d_sd_t}\Delta
	\end{equation*}
	with
	\begin{equation}
	\Delta = \left(t_m'-\dfrac{t_m}{d_m}\right)\dfrac{1}{{d_m}^2}.
	\end{equation}
	\item For compression case $d_n<0$, the components of $\textbf{t}$ are
	\begin{equation}
	t_n = Kd_n, \quad t_s =\dfrac{\beta^2t_m}{d_m}{d_s},\quad t_t =\dfrac{\beta^2t_m}{d_m}{d_t},
	\end{equation}
	so that
	\begin{equation}
	K_{nn} =  K,\quad
	K_{ns} = K_{nt} = 0,
	\end{equation}
	\begin{equation*}
	K_{ss} = \dfrac{\beta^2t_m}{d_m}+\beta^4{d_s}^2\Delta,\quad K_{ss} = \dfrac{\beta^2t_m}{d_m}+\beta^4{d_t}^2\Delta,\quad K_{st} = \beta^4{d_sd_t}\Delta.
	\end{equation*}
\end{enumerate}

The tangent stiffness matrix of the bilinear cohesive law may not be positive definite. At the point of failure, implicit calculations for both DNS and deep material network are unable to converge to an equilibrium solution, which makes it impossible to track the post-failure behavior. Therefore, an extra viscous term is introduced to overcome the convergence problem following the work by Gao et al.\cite{gao2004simple}:
\begin{equation}\label{eq:vis}
\textbf{t} \leftarrow \textbf{t} + \zeta \dfrac{d}{dt}\left(\dfrac{\textbf{d}}{d_f}\right),
\end{equation}
where $\zeta$ is a viscosity-like parameter that governs viscous energy dissipation. By selecting a sufficiently small time increment in the numerical scheme, one is able to make the tangent stiffness tensor of the softening material positive definite.

\bibliography{references_ROM}
\end{document}